\documentclass[12pt]{article}
\newcommand{\beao}{\begin{eqnarray*}}
\newcommand{\eeao}{\end{eqnarray*}}
\newcommand{\be}{\begin{equation}}\newcommand{\ee}{\end{equation}}
\newcommand{\bea}{\begin{eqnarray}}
\newcommand{\eea}{\end{eqnarray}}
\newcommand{\beq}{\begin{eqnarray}}
\newcommand{\eeq}{\end{eqnarray}}
\newcommand{\tra}{\top}
\newcommand{\nn}{\nonumber}
\newcommand{\pa}{\partial}
\newcommand{\ep}{\epsilon}
\newcommand{\al}{\alpha}\newcommand{\la}{\lambda}
\newcommand{\Ref}[1]{(\ref{#1})}
\renewcommand{\ss}{\sinh(s)}\newcommand{\st}{\sinh(t)}
\newcommand{\cs}{\cosh(s)}\newcommand{\ct}{\cosh(t)}

\renewcommand{\sp}{\sinh(s+t)}\newcommand{\sm}{\sinh(s-t)}
\newcommand{\cm}{\cosh(s-t)}

\newcommand{\sq}{\sinh(q)}\newcommand{\cq}{\cosh(q)}
\newcommand{\sxi}{\sinh(\xi)}\newcommand{\cxi}{\cosh(\xi)}
\usepackage{epsfig}
\usepackage{graphicx}
\begin{document}
\title{Neutral gluon polarization tensor in color magnetic background  at finite temperature}
\author{M.~Bordag\thanks{e-mail: Michael.Bordag@itp.uni-leipzig.de}\\
{\small University of Leipzig, Institute for Theoretical Physics,} \\
{\small  Augustusplatz 10/11, 04109 Leipzig, Germany } \\ [12pt]
V. Skalozub\thanks{e-mail: Skalozub@ff.dsu.dp.ua}\\
{\small Dnepropetrovsk National University, 49050 Dnepropetrovsk, Ukraine}}
\date{\small \it Submitted to PRD on Oct 30th}
\maketitle
\begin{abstract}In the framework of SU(2) gluodynamics, we derive the
tensor structure of the neutral gluon polarization tensor in an
Abelian homogeneous magnetic field at finite temperature and
calculate it in one-loop approximation in the Lorentz background
field gauge. The imaginary time formalism and the Schwinger
operator method are used. The latter is extended to the finite
temperature case.  The polarization tensor  turns out to be non
transversal. It can be written as a sum of   ten tensor structures with
corresponding form factors. Seven tensor structures are
transversal, three are not.
 We represent the form factors in terms of double parametric integrals
and the temperature sum which can be computed numerically.
As applications we calculate the Debye mass and the magnetic mass
of neutral gluons in the background field at high temperature. A comparison
with the results of other authors is done.\end{abstract}
%

\section{Introduction}
The investigations  of  QCD  at high temperature carried out in
recent  years have elucidated  the important role of color
magnetic fields.  In Refs. \cite{Cea01},\cite{Cea03} it was
discovered in lattice simulations that sufficiently strong
constant Abelian magnetic fields described by the potential of the
form $A_{\mu}^a = B \delta_{\mu 2} \delta^{a 3} $, where $B$ is
field strength, $a$ is the index of internal symmetry, $\mu$ -
Lorentz index, shift the deconfinement phase transition
temperature $T_c$. In particular, it was  shown that an increase
in the field strength decreases the transition temperature and for
sufficiently strong field strengths $T_c(B)$ can be equal to zero.
On the other hand, in Refs. \cite{Agasian03}, \cite{Demchik06} from
the analysis of  lattice simulations and in
Refs. \cite{Starinets94}, \cite{Bordag00}, \cite{Strelchenko04} from
 perturbative  resummations of daisy graphs in the background field at high
temperature it was found that Abelian chromomagnetic fields of
order $g B \sim g^4 T^2 $, where $g$ is a the gauge coupling, are
spontaneously created.

 These results are of interest not only for
QCD but also for problems of the early universe where strong
magnetic fields of different kind had likely been present
\cite{Pollock03}.  They served as motivations for investigations
began in our recent papers Refs.\cite{Bordag05}, \cite{Bordag06},
which goal is to determine the operator structure of the gluon
polarization tensor in the constant Abelian chromomagnetic background field
at finite temperature. This
 is necessary for investigations of the quark-gluon plasma (QGP),
  first of all, when resummations of perturbative series  are carried
  out. As a preliminary step, the tensor structure of the
gluon polarization tensor as well as  the one-loop contributions
to its form factors  at zero temperature have been obtained and
partially investigated therein. In the  presence of the background field, for many
reasons it is convenient to use the decomposition of the gauge
fields in the internal space of the form $W^{\pm}_{\mu} =
1/\sqrt{2}( A^1 \pm i A^2_{\pm} ), A_{\mu} = A^3_{\mu}$ and
consider  the former "charged" and the latter "neutral" gluons
separately because of sufficiently different properties of them.
That concerns not only physics but also the calculation procedures
required for investigations.

In the present paper in the framework of SU(2) gluodynamics, we
derive the operator structure of the neutral gluon polarization
tensor (PT) in the Abelian chromomagnetic background field at finite
temperature. In actual calculations, as in Refs. \cite{Bordag05},
\cite{Bordag06}, we use the background Lorentz gauge and the
Schwinger operator formalism based on the proper-time
representation for propagators \cite{Schwinger73}. It is modified
in order to account for the Matsubara imaginary time method at
finite temperature and at the same time it preserves the Lorentz
covariance when the momentum loop integrations are fulfilled. This
generalizes the formulas derived earlier at zero temperature. As
applications we calculate the one-loop contribution to the form
factors. For instance, we derive the
Debye mass and the "magnetic mass" of the neutral gluons in the
background field at finite temperature. The limit of zero background  field is
also considered, to the correspondence with known results
obtained already by other methods \cite{Kalashnikov84} is established.

Here to present the results we would like to note that in the
limit $T \to \infty$ the form factors can be calculated in terms
of the Riemann Zeta-function. We also find that the "fictitious"
pole (see for details, for instance, the surveys
\cite{Kalashnikov84}, \cite{Rebhan03}) in the gluon Green's function
at finite temperature disappears in one-loop order if the
external field is switched on. At the same time, the transversal
neutral gluon field modes remain long range ones. They are not
screened in the field at one-loop order, in contrast to the
charged ones, as it was determined in Refs. \cite{Strelchenko00},
\cite{Strelchenko04}. This difference is important. Its possible
consequences  will be discussed below in the main text.

The paper is organized as follows.  In the next section we
introduce the necessary notation and review in brief the basic
formulas. In sections 3, 4 we derive the operator structure of the
gluon polarization tensor and develop the calculation procedure to
carry out the integration over  internal momenta  of Feynman
diagrams in the field at finite temperature. Explicit formulas for
the form factors in the form of two-parametric integrals are
obtained in section 5. In contrast to the previous paper
\cite{Bordag05}, where the tadpole diagrams were not discussed in
detail and instead some arguments in fewer of the cancellation of
this contributions in the total  by the surface terms appearing
in other parts of the polarization tensor were used.  Here we consider
 this cancelation because of peculiarities appearing
  at finite temperature. In section 6 the transition to
the zero temperature case is discussed. The Debye mass of the
neutral gluon in the external field is calculated in section 7. In the
next section we calculate the mean values of the operator in the
physical states of the transverse modes and show that the gluon
"magnetic mass" in the field is zero in one-loop order although
the fictitious pole of the Green function is eliminated. The
discussion of the results obtained and further prospects are given
in section 9.

Throughout the paper we use latin letters $a,b,\dots=1,2,3$ for
the color indexes and Greek letters $\la,\mu,\dots=1,\dots,4$ for
the Lorentz indices.  Summation over doubly appearing indices is
assumed. All formulas are in the Euclidean formulation. We put all
constants including the coupling equal to unity. Since the present
work is a continuation of investigations began in
Ref.\cite{Bordag05}, we follow the notations, definitions and
calculation procedures used therein as close as possible.
\section{Basic Notations}\label{bn}
In this section we collect the well known basic formulas for SU(2)
gluodynamics to set up the notations which we will use.
  We work in the Euclidean version. Dropping
arguments and indices, the Lagrangian is
\be\label{BL}{\cal L}=-\frac14 F_{\mu\nu}^2-\frac{1}{2\xi}(\pa_\mu
A_\mu)^2+\overline{\eta} \ \pa_\mu D_\mu \eta, \ee
where $\xi$ is the gauge fixing parameter and $\eta$ is the ghost
field. The action is $S=\int dx \ L$ and the generating functional
of the Green functions is $Z=\int DA \exp(S)$. In the following we
divide the gauge field $A_\mu^a(x)$ into background field
$B_\mu^a(x)$ and quantum fluctuations $Q_\mu^a(x)$,
\be\label{ABQ}A_\mu^a(x)=B_\mu^a(x)+Q_\mu^a(x). \ee
The covariant derivative depending on a field $A$ is
\be\label{covder}D_\mu^{ab}[A]=\frac{\pa}{\pa
x^\mu}\delta^{ab}+\ep^{acb}A_\mu^c(x) \ee
and the field strength is
\be\label{fs}F_{\mu\nu}^a[A]=\frac{\pa}{\pa
x^\mu}A_\nu^a(x)-\frac{\pa}{\pa
x^\nu}A_\mu^a(x)+\ep^{abc}A_\mu^b(x)A_\nu^c(x) \ee
and
\be\label{komm}\left[D_\mu[A],D_\nu[A]\right]^{ab}=\ep^{acb}F_{\mu\nu}^c[A]
\ee
holds. For the field splited  into background and quantum parts we
note
\bea\label{F}F_{\mu\nu}^a[B+Q]&=&F_{\mu\nu}^a[B]+D_\mu^{ab}[B]Q_\nu^b(x)
-D_\nu^{ab}[B]Q_\mu^b(x) \nn \\ && +\ep^{abc}Q_\mu^b(x)Q_\nu^c(x).
\eea
The square of it is
\bea\label{FBQ}-\frac14 \left(F_{\mu\nu}^a[B+Q]\right)^2&=&
-\frac14
\left(F_{\mu\nu}^a[B]\right)^2+Q_\nu^aD_\mu^{ab}[B]F_{\mu\nu}^b[B]\nn
\\ && -\frac12 Q_\mu^a K_{\mu\nu}^{ab}Q_\nu^b +{\cal M}_3+{\cal
M}_4. \eea
The second term in the r.h.s. is linear in the quantum field and
disappears if the background fulfills its equation of motion which
will hold in our case of a constant background field. The third term is
quadratic in $Q_\mu^a$ and it defines the 'free part' with the
kernel
\be\label{K}K_{\mu\nu}^{ab}=-
\delta_{\mu\nu}D_\rho^{ac}[B]D_\rho^{cb}[B]+D_\mu^{ac}[B]
D_\nu^{cb}[B] -2\ep^{acb}F_{\mu\nu}^c[B]. \ee
The interaction of the quantum field  is represented by the
vertex factors
\bea\label{vfM}{\cal M}_3&=&-\ep^{abc}\left(D_\mu^{ad}Q_\nu^d\right)Q_\mu^bQ_\nu^c ,\nn \\
{\cal M}_4&=&-\frac14 Q_{\mu}^{a}Q_{\nu}^{a}Q_{\mu}^{b}Q_{\nu}^{b}
+ \frac14 Q_{\mu}^{a}Q_{\nu}^{b}Q_{\mu}^{a}Q_{\nu}^{b} . \eea
The complete Lagrangian
\be\label{comLa}{\cal L}=-\frac14
\left(F_{\mu\nu}^a[B+Q]\right)^2+{\cal L}_{\rm gf}+{\cal L}_{\rm
gh} \ee
consists of \Ref{FBQ}, the gauge fixing term (in the following we
put $\xi=1$),
\be\label{Lgf}{\cal L}_{\rm gf}=-\frac{1}{2\xi}\left(D_\mu^a[B]
Q_\mu^a\right)^2= \frac{1}{2\xi}Q_\mu^a
D_\mu^{ac}[B]D_\nu^{cb}[B]Q_\nu^b, \ee
and the ghost term
\be\label{Lgh}{\cal L}_{\rm
gh}=\overline{\eta}^aD_\mu^{ac}[B]\left(D_\mu^{cb}[B]+\ep^{cdb}Q_\mu^d\right)\eta^b.
\ee
These formulas are valid for an arbitrary background field.  Now
we turn to the specific background of an Abelian homogeneous
magnetic field of strength $B$ which is oriented along the third
axis in both, color and configuration space. An explicit
representation of its vector potential is
\be\label{Bexpl}B_\mu^a(x)=\delta^{a3}\delta_{\mu 1}x_2 \ B \ee
and the corresponding field strength is
\be\label{}F_{ij}^a=\delta^{a3}F_{ij}=B\ep^{3ij},\ee
where only the spatial components ($i,j=1,2,3$) are non\-zero.
Once the background is chosen Abelian it is useful to turn into
the so called charged basis,
\bea\label{}W^\pm_\mu&=&\frac{1}{\sqrt{2}}\left(Q_\mu^1\pm i Q_\mu^2\right) \nn \\
Q_\mu&=&Q_\mu^3 \eea
with the interpretation of $W^\pm_\mu$ as color charged fields and
$Q_\mu$ as color neutral field. This is in parallel to
electrically charged and neutral fields. Note also that $Q_\mu$ is
real while $W^\pm_\mu$ are complex conjugated one to the other. In
the following we will omit the word color when speaking about
charged and neutral objects. The same transformation is done for
the ghosts,
\bea\label{}\eta^\pm_\mu&=&\frac{1}{\sqrt{2}}\left(\eta_\mu^1\pm i \eta_\mu^2\right) ,\nn \\
\eta_\mu&=&\eta_\mu^3. \eea
A summation over the color indices turns into
\be\label{}Q^a_\mu Q^a_\nu=Q_\mu Q_\nu+W^+_\mu W^-_\nu+W^-_\mu
W^+_\nu . \ee
All appearing quantities have to be transformed into that basis.
For the covariant derivative we obtain
\be\label{}\begin{array}{rclrclll}D^{33}_\mu&=&\pa_\mu,&
D^{-+}_\mu&=&\pa_\mu-iB_\mu&\equiv& D_\mu \\ [8pt] &&&
D^{+-}_\mu&=&\pa_\mu+iB_\mu  &\equiv& D^*_\mu \end{array} \ee
where $D^*_\mu$ is the complex conjugated to $D_\mu$. Starting
from here we do not need any longer to indicate the arguments in
the covariant derivatives.

Before proceeding with writing down the remaining formulas in the
charged basis it is useful to turn into momentum representation.
This can be done in a standard way by the formal rules. It remains
to define  the signs in the exponential factors. We adopt the
notation
\be\label{}Q\sim e^{-ikx}, \qquad W^-\sim e^{-ipx}, \qquad W^+\sim
e^{ip'x}. \ee
In all following calculation the momentum $k$ will denote the
momentum of a neutral line and the momenta $p$ and $p'$ that of
the charged lines whereby $k$ and $p$ are incoming and $p'$ is
outgoing. In these notations the covariant derivative $D_\mu$ acts
on a $W^-_\mu$ and turns into
\be\label{defp}D_\mu=-i(i\pa_\mu+B_\mu)\equiv -i p_\mu . \ee
Note that the components of the momentum $p_\mu$ do not commute,
\be\label{noncp}[p_\mu,p_\nu]=iBF_{\mu\nu}. \ee
In these notations the quadratic term of the action turns into
\bea&&\label{q2} -\frac12 Q_\mu^a K_{\mu\nu}^{ab}Q_\nu^b= \\ &&
\frac12 Q_\mu K_{\mu\nu}^{33} Q_\nu+\frac12 W^+_\mu
K_{\mu\nu}^{-+}W^-_\nu+ \frac12 W^-_\mu K_{\mu\nu}^{+-}W^+_\nu \nn
\eea
with
\be\label{Kk} K_{\mu\nu}^{33}\equiv
K_{\mu\nu}(k)=\delta_{\mu\nu}k^2-k_\mu k_\nu \ee
and
\be\label{Kp}K_{\mu\nu}^{-+}\equiv
K_{\mu\nu}(p)=\delta_{\mu\nu}p^2-p_\mu p_\nu+2iBF_{\mu\nu}. \ee
We use the arguments $k$ and $p$ instead of the indices to
indicate to which line a $K_{\mu\nu}$ belongs. The third term in
the r.h.s. of Eq.\Ref{q2} is the same as the second one due to the
complex conjugation rules. In the Feynman rules
$(K_{\mu\nu}^{33})^{-1}$ is the line for neutral gluons and is
denoted by a wavy line and $(K_{\mu\nu}^{-+})^{-1}$ is the line
for charged gluons and is denoted by a directed solid line. We
remark that these lines represent propagators in the background of
the magnetic field. Frequently they are denoted by thick or double
lines. Because we have in this paper no other lines the notation
with ordinary (thin) lines is unique.

Here we note that the spectrum of the operator \Ref{Kp} in a
constant magnetic field,
 \be\label{spectr} E_n^2= p^2_3 + B(2n+1),~ n=-1,0,1,... , \ee
contains a tachyonic mode at $n=-1.  ~p_3 $ is momentum along the
field direction $B = B_3$. This state is a peculiar of non Abelian
gauge fields as it is discussed in different aspects in the
literature (see, for instance Ref.\cite{Bordag05},\cite{Bordag00}
and references therein).

 For later use we introduce here the set of eigenstates for
the operator \Ref{Kk}. For the color neutral states we take exactly
the same polarizations $\mid k,s>$ as known from electrodynamics,
\be\label{qedstates}
\begin{array}{rclrcl}
\mid k,1>_\mu&=&\frac{1}{h}\left(\begin{array}{c}
-k_2\\k_1\\0\\0\end{array}\right)_\mu,&
\mid k,2>_\mu&=&\frac{1}{k h}\left(\begin{array}{c} k_1 k_3\\k_2 k_3\\-h^2\\0\end{array}\right)_\mu , \\
\mid k,3>_\mu&=&\frac{1}{k}\left(\begin{array}{c}
k_1\\k_2\\k_3\\0\end{array}\right)_\mu,& \mid
k,4>_\mu&=&\left(\begin{array}{c} 0\\0\\0\\1\end{array}\right)_\mu
\end{array}
\ee
with $h =\sqrt{k_1^2+k_2^2}$, $k=\sqrt{k_1^2+k_2^2+k_3^2}$. Here
the polarizations $s=1,2$ describe the two transversal gluons
($k_\mu\mid k,s=1,2>_\mu=0$), $s=3$ is the longitudinal one and
$s=4$ after rotation into Minkowski space becomes the time like
one. For the transversal gluons
\be\label{eomK} K_{\mu\nu}(k)  \ \mid k,s=1,2>_\nu=(k_4^2+k^2) \
\mid k,s=1,2>_\mu \ee
holds.

\begin{figure}
\begin{picture}(10,5)
\put(5,0){\psfig{height=4cm,file=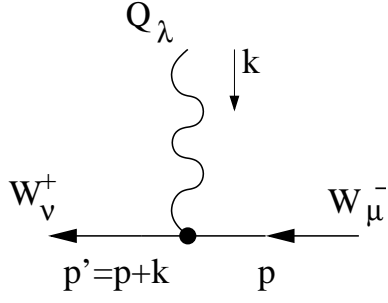}}
\end{picture}
\caption{Notations for the vertex ${\cal M}_3$}
\label{figure:Gamma3}
\end{figure}

After discussing the free part of the Lagrangian \Ref{comLa},
\Ref{FBQ},  in the 'charged basis' we note that for the vertex
factor ${\cal M}_3$ in  \Ref{vfM} we obtain
\be\label{M3}{\cal M}_3=W^-_\mu\Gamma_{\mu\nu\la}W^+_\nu Q_\la \ee
with
\be\label{Gamma}\Gamma_{\mu\nu\la}=\delta_{\mu\nu}(k-2p)_\la+\delta_{\la\mu}(p+k)_\nu+\delta_{\la\nu}(p-2k)_\mu
. \ee
The notations are shown in Fig.\ref{figure:Gamma3}. It should be
remarked that all graphs and combinatorial factors are exactly the
same as in the well known case without magnetic field. On this
level the only difference is in the meaning of the momentum
$p_\mu$ which   in our case depends on the background magnetic
field, see Eq.\Ref{defp}.

\unitlength1cm

%
%

The vertex factor describing the interaction of the  neutral
gluons with charged ghost fields is
\be\label{M3}{\cal M}^{gh.}_3=\eta^- \Gamma_{\la}^{q \eta} \eta^+
Q_\la \ee
where
\be\label{Gammagh} \Gamma_{\la}^{q \eta}  = p'_\la = (p + k)_\la .
\ee
We also need in the four particle vertexes which are momentum
independent and have the same form as at zero external field.
\section{Operator Structures}\label{sectopstr} The neutral PT is
denoted by $\Pi_{\la\la'}(k)$ where the argument $k$ is an
ordinary momentum. Its one-loop diagram representation is shown in
Fig. \ref{figure:Pineu}.
 In this section we discuss the general tensor
structure of it at zero and finite temperature. As it was shown in
Refs. \cite{Bordag05},\cite{Bordag06} it  is not transversal in a
magnetic background field. That means that the condition
$k_\la\Pi_{\la\la'}(k)=0$ does not hold. This follows either from
the Slavnov-Taylor identity for the gluon Green function or from
an explicit one-loop calculation. So,  we are left with the weaker
condition
\be\label{doubletransv} k_\la \ \Pi_{\la\la'}(k) \ k_{\la'}=0 .
\ee

At zero temperature, it can be combined with the remaining Lorentz
symmetry which results in  a dependence of $\Pi_{\la\la'}(k)$ on
two vectors, $l_\la$ and $h_\la$, and on the magnetic field.

We use the notations \bea&&\label{vec}
l_\mu=\left(\begin{array}{c}0\\0\\k_3\\k_4\end{array}\right),~
h_\mu=\left(\begin{array}{c}k_1\\k_2\\0\\0\end{array}\right), ~
d_\mu=\left(\begin{array}{c}k_2\\-k_1\\0\\0\end{array}\right),~
\nn\\&&
F_{\mu\lambda}=\left(\begin{array}{cccc}0&1&0&0\\-1&0&0&0\\0&0&0&0\\0&0&0&0\end{array}
\right).\eea

The third vector is $d_\mu\equiv F_{\mu\nu}k_\nu$. Note that here
and further  in actual calculations the magnetic field strength
$B$ is put equal to unity. For the vectors $k_\la=l_\la+h_\la$
holds.

The general structure of $\Pi_{\la\la'}(k)$ at $T = 0$ allowed by
\Ref{doubletransv} and the vectors $l_\la$ and $h_\la$ is
determined by the set of tensor structures
\begin{eqnarray} \label{Tn}
T^{(1)}_{\la\la'}&=& l^2 \delta^{||}_{\la\la'}-l_\la l_{\la'}\nn \\
T^{(2)}_{\la\la'}&=&h^2\delta^\perp_{\la\la'}
-h_\la h_{\la'}=d_\la d_{\la'}    \nn \\
T^{(3)}_{\la\la'}&=& h^2 \delta^{||}_{\la\la'}+
l^2 \delta^\perp_{\la\la'} -l_\la h_{\la'}-h_\la l_{\la'}\nn \\
T^{(4)}_{\la\la'}&=& i(l_\la d_{\la'}-d_\la l_{\la'} )
+il^2F_{\la\la'}\nn \\
T^{(5)}_{\la\la'}&=&h^2 \delta^{||}_{\la\la'}-l^2 \delta^\perp_{\la\la'} \nn \\
T^{(6)}_{\la\la'}&=&iF_{\la\la'}
\end{eqnarray}
together with the identity
$d_{\la}h_{\la'}-h_{\la}d_{\la'}=h^2F_{\la\la'}$. Further  we
introduced the notations
$\delta^\perp_{\mu\lambda}=\mbox{diag}(1,1,0,0)$ and
$\delta^{||}_{\mu\lambda}=\mbox{diag}(0,0,1,1)$. The first four
operators are transversal, $k_\la
T^{(i)}_{\la\la'}=T^{(i)}_{\la\la'}k_{\la'}=0$ with $i=1,2,3,4$,
the last two fulfill  \Ref{doubletransv}, only. The sum of the
first three operators is just the transversal part of the kernel
of the quadratic part of the action, Eq. \Ref{Kk},
\be\label{sumTip}
T^{(1)}_{\la\la'}+T^{(2)}_{\la\la'}+T^{(3)}_{\la\la'}=K_{\la\la'}(k).
\ee
At finite temperature, an additional vector $u_{\mu}$ - the
velocity of the thermostat - must be accounted for and used in
construction of the tensors $T^{(i)}$. Therefore new tensor
structures appear. We chose  them in the form:
\begin{eqnarray} \label{TnT}
T^{(7)}_{\la\la'}&=& (u_\la l_{\la'} + l_\la u_{\la'})(u k) - \delta^{||}_{\la \la'}(u k)^2 - u_\la u_{\la'} l^2 \nn \\
T^{(8)}_{\la\la'}&=& (u_\la h_{\la'} + h_\la u_{\la'})(u k) - \delta^{\perp}_{\la\la'}(u k)^2 - u_\la u_{\la'} h^2 \nn \\
T^{(9)}_{\la\la'}&=& u_\la i d_{\la'} - i d_\la u_{\la'} + i
F_{\la\la'} (u k),  \nn \\
T^{(10)}_{\la\la'}&=& k^2 \delta_{\la\la'} - \frac{u_\la
u_{\la'}(k^2)^2}{(u k)^2}.
\end{eqnarray}
Obviously the sum \be\label{sum78} T^{(7)}_{\la\la'}+
T^{(8)}_{\la\la'}= (u_\la k_{\la'} + k_\la u_{\la'})(u k) -
\delta_{\la\la'}(u k)^2 - u_\la u_{\la'} k^2 = B_{\la\la'} \ee
 equals to  one of two transversal tensor structures commonly used at
  zero field \cite{Kalashnikov84}. In that case the first structure  is
given by the sum of tensors Eqs. \Ref{sumTip}, \Ref{Kk}. Below in
actual calculations we use the reference frame of the thermostat,
so only one component of $u_{\mu}$ is nonzero: $u_{\mu}=
(0,0,0,1)_\mu$.
We mention that the first two tensors in Eq.\Ref{TnT} are
 transversal and the other two satisfy the weaker condition
 \Ref{doubletransv}.

 We adopt the following way for the
 representation of our expressions. The dimensionality of the
 polarization tensor is
 implemented in the tensors $T^{(i)}$. To restore the  
  dimensionality for the tensors in Eqs.\Ref{Tn} and \Ref{TnT},
   one has to multiply the operator
  $T^{(6)}$  by the factor $B$, and the operator
  $T^{(9)}$  by  $\sqrt{B}.$
  The form factors are  dimension less functions of dimension
  less momenta $l^2, h^2,$ 
   and temperature $ T$. That means, in fact they are measured in units
  of $B$. To restore the
correct dimensionality one has to replace $l^2 \to l^2/B, h^2 \to
h^2/B$, and $T\to T/\sqrt{B}$. Correspondingly, the arguments of all functions 
appearing in the actual calculations are also dimension less.

 Knowing the operators \Ref{Tn} and
\Ref{TnT},  which may appear at zero and finite temperature, the
polarization tensor can be represented in the form
\be\label{exp} \Pi_{\la\la'}(k)=\sum_{i=1}^{10} \ \Pi^{(i)}(k) \
T^{(i)}_{\la\la'} \ ,\ee
where the form factors $\Pi^{(i)}(k)$ depend on the external momentum
$k_\mu$  through the variables $l^2$ and $h^2$ at zero temperature
and $h^2, k_4$ and $k_3$ at finite temperature. In the former
case, the polarization tensor $\Pi_{\la\la'}(k)$ is real and
symmetric in its indices, so the form factors $\Pi^{(4)}(k)$ and
$\Pi^{(6)}(k)$ are zero. In the latter case all structures will
contribute in general.

In the Matsubara formalism,  it is possible to add to the set
of tensors Eq.\Ref{TnT} a structure of a special type which
contributes in the static limit, $k_4 = 0 $, only (remember, $k_4$ takes discrete values). It should be of
the form $ u_\mu u_\nu \phi(k_4)$ with $\phi(k_4)$ possesses the
following property: it is non zero for $k_4 = 0$, only.  This term
is obviously transversal for itself. So, the set of tensors in
Eq.\Ref{exp} could be extended due to this structure. On the other side this structure is a linear combination of $T^{(7)}_{\la\la'}$ and $T^{(8)}_{\la\la'}$ at $k_4=0$.

\section{Calculation of the Neutral Polarization
Tensor}

 In this section we calculate the one-loop contribution to
the PT at finite temperature. The imaginary time formalism is
used. That means that in loops the integration over momentum
component $p_4$ is substituted by an infinite series in discrete
values $ p_4 = 2 \pi N T$: $\int\limits_{- \infty}^{+ \infty}
\frac{d p_4}{2 \pi} f(p_4)\to T \sum\limits\limits_{N = -
\infty}^{ + \infty}f(2 \pi N T)$.

\begin{figure}[htbp]\begin{picture}(10,3.9)
\put(1,0.19){ \psfig{height=4cm,file=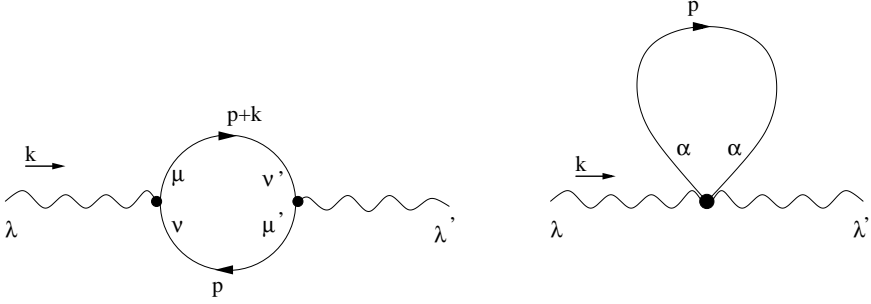}}
\end{picture}
\caption{The neutral polarization tensor} \label{figure:Pineu}
\end{figure}

 The neutral polarization tensor has the following
representation in momentum space  (see Fig. \ref{figure:Pineu})
\be\label{NPT}\Pi_{\la\la'}(k)= T \sum\limits_{N=- \infty}^{+
\infty}\int\limits_{- \infty}^{+ \infty}\frac{d^3 p}{(2
\pi)^3}~\Pi(p,p_4,k,k_4)_{\la\la'},\ee
where in the integrand we noted explicitly the dependence  on the
external momentum and the momentum inside loops. In what follows,
as in Ref.\cite{Bordag05}, at intermediate stage of computation
we, for brevity, shall omit the general factors and the signs of
integration and summation. That is, we relate these factors and
operations with the momentum $p$ standing in r.h.s. of equations.
Within this convention we write
 \bea\label{NPT1}
\Pi_{\la\la'}(k)&=&\Gamma_{\mu\nu\la}G_{\mu\mu'}(p)\Gamma_{\mu'\nu'\la'}
G_{\nu'\nu}(p-k)\\&&
-p_{\la}G(p)(p-k)_{\la'}G(p-k)-(p-k)_{\la}G(p)p_{\la'}G(p-k)\nn \\
&& + G_{\la\la'}(p) + G_{\la'\la}(p) - 2 \delta_{\la\la'} Tr
G(p).\nn \eea
  The second line gives
the contribution from the ghost loops and the third one is due to
the tadpole diagram.

The vertex factor is given in Eq.\Ref{Gamma}.
 For a convenient grouping of terms it
is useful to rearrange it,
\be\label{vertexfactor} \begin{array}{lcccccc}
\Gamma_{\mu\nu\la}= \\
\underbrace{g_{\mu\nu}(k-2p)_{\la}}&+&
\underbrace{2\left(g_{\la\mu}k_\nu-g_{\la\nu}k_\mu\right)}&+&
\underbrace{g_{\la\mu}(p-k)_\nu+g_{\la\nu}p_\mu} , \\ [10pt]\equiv
\Gamma_{\mu\nu\la}^{(1)}&+&\Gamma_{\mu\nu\la}^{(2)}&+&\Gamma_{\mu\nu\la}^{(3)},
\end{array}
\ee
where in the last line a subdivision into three parts is done.

The propagators are given by
\bea\label{prop0} &&G(p)=\frac{1}{p^2}=\int_0^\infty ds \
e^{-sp^2}, \nn \\ && G(p-k)=\frac{1}{(p-k)^2}=\int_0^\infty dt \
e^{-t(p-k)^2}  \eea
 for the scalar lines
and by \bea\label{prop} G_{\la \la'}
(p)&=&\left(\frac{1}{p^2+2iF}\right)_{\la\la'}=\int_0^\infty ds \
e^{-sp^2}E^{-s}_{\la\la'}  ,\nn\\
G_{\la \la'}
(p-k)&=&\left(\frac{1}{(p-k)^2+2iF}\right)_{\la\la'}\\&&~~~~~~~~~=\int_0^\infty
dt \ e^{-t(p-k)^2}E^{-t}_{\la\la'} \nn \eea for the vector lines
(in the Feynman gauge, $\xi = 1)$ with
\bea\label{E}E^s_{\la\la'} &\equiv&
\left(e^{2isF}\right)_{\la\la'}  \nn \\&=&
\delta^{||}_{\la\la'}+iF_{\la\la'}\sinh(2s)+\delta^\perp_{\la\la'}
\cosh(2s) . \eea

At zero temperature, the momentum integration can be carried out
by means of Schwinger's  algebraic procedure \cite{Schwinger73}
and converted into an integration over two scalar para\-meters,
$s$ and $t$. Here we educe the known results in order to present
their modifications at $T \not = 0$.  The basic exponential is
\be\label{mexp} \Theta=e^{-sp^2}e^{-t(p-k)^2}\ee
and the  integration over the momentum $p$ is denoted by the
average $\langle\dots\rangle$.
 The following formulas hold:
\bea \label{Theta}\langle\Theta\rangle= \frac{\exp
\left[-k\left(\frac{st}{s+t}\delta^{||}+\frac{ST}{S+T}\delta^{\perp}\right)
k \right]}{(4\pi)^2(s+t)\sinh(s+t)} \eea with $S=\tanh(s)$ and
$T=\tanh(t)$ and
\be\label{av1}   \langle p_\mu \ \Theta\rangle  =
\left(\frac{A}{D}k\right)_\mu \langle \Theta\rangle , \ee
\be\label{av2}   \langle p_\mu p_\nu \ \Theta\rangle  =
\left(\left(\frac{A}{D}k\right)_\mu \left(\frac{A}{D}k\right)_\nu
-i \left(\frac{F}{D^\tra}\right)_{\mu\nu}\right)\langle
\Theta\rangle .\ee
The notation $A\equiv E^t-1$ and $D\equiv E^{s+t}-1 $ is used.
Explicit formulas are
\be\label{A/D}  \frac{A}{D}=\delta^{||}\frac{t}{s+t}
-iF\frac{\sinh(s)\sinh(t)}{\sinh(s+t)}
+\delta^\perp\frac{\cosh(s)\sinh(t)}{\sinh(s+t)} \ee
along with
\be\label{FED}\frac{-2iFE^{-s}}{D^{\top}}=\frac{\delta_{||}}{s+t}
-i F \frac{\sm}{\sp}+\delta_\perp \frac{\cm}{\sp}, \ee
where we dropped the indices. It should be remarked that all these
matrices, i.e., $E$, $F$, $D$ and $A$ commute. In addition we need the
relation
$$ p(s)_\mu\equiv e^{-sp^2}p_\mu e^{sp^2}={E^s}_{\mu\nu} p_\nu $$
for commuting a factor $p_\mu$ with the propagator $G(p)$,
\be\label{cp0} p_{\mu}G_{\mu\mu'}(p)=G(p)p_{\mu'}. \ee

Now we turn to the finite temperature case. Our goal is to account
for the temperature dependence within the above
representation  in a natural way. Usually in the imaginary time formalism the
summation over $p_4$ and the integration over three-momenta are carried
out separately. To restore the equivalence of these variables and to
make use of the formulas \Ref{Theta}-\Ref{av2} we proceed in the
following way. First we note that any function $ f(p_4 = 2 \pi N
T)$ of $p_4$ can be written in the form
\bea \label{intf} f(p_4=2\pi N T)&=& \int d p_4 f(p_4) \delta(
p_4 - 2 \pi N T)\nn \\
& =& \int d p_4 f(p_4) \frac{1}{2 \pi} \int\limits_{- \infty}^{+
\infty}d \la e^{i \la (p_4 - 2 \pi N T)}, \eea %
where $\delta(x)$ is Dirac's delta-function. Then we change the
order of integration in variables $\la$ and $p_4$ and fulfill the
momentum integration as at zero temperature. The factor $1/(2\pi)$
coming from the delta-function and the factor $1/(2\pi)^3$ coming from the
three-momentum integration give in the product the  factor
$1/(2 \pi)^4$ appearing at zero temperature. The only new factor,
$e^{i \la p_4} $, appears as another exponential in Eq.\Ref{mexp}. Since
the fourth component is not related with the magnetic field, the
calculation is actually the same as at zero temperature. Remind
that in course of it the factor $1/(2\pi)^4$ results in the factor
$1/(4\pi)^2$ in the function $\langle \Theta (s, t) \rangle$
\cite{Schwinger73}.  The integral over $p_4$ is Gaussian,
 and we obtain for the algebraic averaging (or bracket) procedure
$\langle\dots\rangle$ at $\la \not = 0$
\be\label{ThetaTla} \langle \Theta (s, t, \la) \rangle_\la =
\langle e^{i \la p_4} \Theta(s,t) \rangle = \langle
\Theta(s,t)\rangle \ \exp\left(- \frac{\la^2}{4 q} + i \la k_4
\frac{t}{q}\right), \ee
where  we  marked the averaging procedure with $\la$-dependence
 by the subscript $\la$. Here $k_4$ is the
discrete fourth component of the  external momentum and we introduced $q = s + t$ as a
convenient variable. The expression in the angle  brackets in the
r.h.s. is the zero temperature value given in Eq.\Ref{Theta}.
Below we will denote the function \Ref{ThetaTla} as
$\Theta(s,t,\la)$.

At the next step, we  integrate the function $ \Theta (s, t, \la)$
over $\la$ and after the restoration of the sum over $N$ we obtain the
basic  expression at finite temperature:
\bea \label{ThetaT} &&\langle\Theta(s,t)\rangle_T= T \sum\limits_{
N = -\infty}^{ +\infty} \int\limits_{-\infty}^{+\infty} d\la
\Theta(s, t, \la) e^{ - i \la (2\pi N T)} \nn\\
&&= T \sum\limits_{N = -\infty}^{ +\infty}  \sqrt{4\pi q}\langle
\Theta (s, t) \rangle \ \exp\left( - k_4^2 \frac{t^2}{q} + 2 k_4 t (2 \pi
N T) - (2 \pi N T)^2 q \right). \eea
We denote the function standing under the sign $\sum$ by
$\Theta_T$.

To obtain the result of
 the bracket procedure with the momentum $p_4$ entering, one
has to differentiate  Eq.\Ref{ThetaTla} with respect to  $i\la$
and then to calculate the integral over $\la$. This  can be done
also by means of differentiation of Eq.\Ref{ThetaT} with respect
to the parameter $b_N \equiv 2 \pi N T$. In this way the all integrals
of interest can be computed.

The expressions with the spatial indexes $i, j = 1,2,3$ are given
by Eqs.\Ref{Theta}-\Ref{av2}, where the function \Ref{ThetaT} must
be substituted.

The average with one spatial component and $p_4$ is
\bea \label{pp4} \langle p_i p_4\Theta (s, t)\rangle_T&=&T
\sum\limits_{ N = -\infty}^{ +\infty}
\int\limits_{-\infty}^{+\infty} d\la  e^{ - i \la (2\pi N
T)}\frac{k_4 + i \la/2}{q} \left(\frac{A}{D} k\right)_i \Theta (s,
t, \la)\nn\\
&&= \sum\limits_{ N = -\infty}^{ +\infty} \left(\frac{A}{D}
k\right)_i \ 2\pi N T \ \Theta (s, t, \la). \eea
The bracket procedure with $p_4^2$ results in the following
expression:
\bea \label{pp42} \langle p_4^2\Theta(s, t) \rangle_T&=&T
\sum\limits_{ N = -\infty}^{ +\infty}
\int\limits_{-\infty}^{+\infty} d\la \  e^{ - i \la (2\pi N
T)} \ \left( \left( \frac{k_4 + i \la/2}{q}\right)^2 + \frac{1}{2 q}
\right)\Theta(s, t, \la)
\nn\\
&&= \sum\limits_{ N = -\infty}^{ +\infty}\Theta_T (s, t) \ ( 2\pi N
T )^2 . \eea
 Note as an interesting fact
that the $k_4$-dependence in the last two equations comes in through
the exponential, only.

 Thus, we  collected the necessary integrals over internal
momentum which appear in the magnetic background field at finite temperature.
These expressions are useful when the high temperature limit is
investigated because due to the exponential factor in
Eq.\Ref{ThetaT} a few first terms in the series $(N=0,1,2,...)$ give the
leading contributions at $T \to \infty$.  The factor $ \sqrt{q}$
entering the integrals over $s$ and $t$ parameters ensures the
convergence at $s,t\to 0$. This corresponds to the superficial divergence degree. In the four dimensional theory the form factors have  zero superficial divergence degree, hence the three dimensional theory which effectively appears in the high temperature expansion is ultraviolet finite.
The convergence at infinity is due to the
exponential factors except for the complications resulting from the tachyonic mode which will be discussed later. However, the  low temperature limit is less trivial.

Now we consider another representation for the integrals,
convenient at low temperature and  carry our the ultraviolet
renormalization. For this purpose we make   resummations of the
form
\be\label{resum} \sum\limits_{ N = -\infty}^{ +\infty} e^{(- z N^2
+ a N)} = \sqrt{\frac{\pi}{z}} \sum\limits_{ N = -\infty}^{
+\infty} e^{\frac{(a + 2 \pi i N)^2}{4 z}} \ee
 of the series in $N$ in the Eq.\Ref{ThetaT} with the parameters $
 a = 4\pi T k_4 t$ and $z = 4\pi^2 T^2 q$ chosen. In this case the
 dependence on $k_4^2$ in the exponential disappears and we obtain
for $\langle\Theta (s, t)\rangle_T$:
\be \label{ThetaT1} \langle\Theta (s, t)\rangle_T = \sum\limits_{N
= -\infty}^{ +\infty} \langle \Theta (s, t) \rangle \ \ \exp\left(-
\frac{N^2}{4 T^2 q} + i \frac{k_4 t N}{q T} \right). \ee
This is just the representation of interest. As above, the
function standing under the sign of the sum will be denoted by
$\Theta_T$.

To obtain the functions $\langle\Theta (s, t) p_4\rangle_T$ and
$\langle\Theta(s, t) p_4^2 \rangle_T$ we have to calculate the
integrals over the variable $\la$ with the powers of $\la$ and $
\la^2$ in the Eqs.\Ref{pp4} and \Ref{pp42} by means of
differentiation of the expression \Ref{ThetaT1} with respect to
the parameter $b_k = ik_4 t/q$ one and two times, correspondingly.
In this way we obtain in the first case,
\be\label{p4res} \langle p_4\Theta (s, t) \rangle_T =
\sum\limits_{ N = -\infty}^{ +\infty} \
\left(  k_4 \frac{t}{q} + \frac{i
N}{2 q T} \right) \ \langle\Theta (s, t)\rangle \
\exp\left(-\frac{N^2}{4 q T^2} +
\frac{i k_4 N t}{q T} \right). \ee
This expression can be combined with the bracket operation for the
spatial momenta $p_i$. In fact, the function $(\frac{A}{D} k)_{\mu}$
for $\mu = 4 $ equals to $ (t/q)k_4$ that coincides with the
first term in the bracket in the above equation. To account for
the second term, we assume that the thermostat is at rest and
therefore the vector $u_{\mu}$ has only one non zero component,
$u_\mu = \delta_{\mu 4}$. To write down Eq.\Ref{p4res} in short,
we introduce the notation $\tilde{u}_\mu = \frac{i N}{2 q T}
u_\mu$. Then, the average operation with one internal momentum
yields
\be\label{pres} \langle p_\mu\Theta (s, t)\rangle_T =
\sum\limits_{ N = -\infty}^{ +\infty}(  (\frac{A}{D}k)_{\mu} +
\tilde{u}_\mu )\Theta (s, t)_T . \ee
In the same manner, in case of two internal momenta the result can
be presented  in the form,
\be\label{p2res}\langle p_\mu p_\nu\Theta(s,t) \rangle_T=
\sum\limits_{ N = -\infty}^{ +\infty}
\left[\left(\left(\frac{A}{D}k\right)_{\mu} +
\tilde{u}_\mu \right)\left(\left(\frac{A}{D}k\right)_{\nu} + \tilde{u}_\nu\right) - i
\left(\frac{F}{D^\tra}\right)_{\mu\nu}\right]\Theta _T(s, t). \ee
Thus, we have calculated  all integrals needed for what
follows.

The derived expressions \Ref{Theta}-\Ref{av2} and
\Ref{ThetaT1}, \Ref{pres}, \Ref{p2res} solve the problem on the
momentum integration in the Schwinger operator formalism at finite
temperature. The results are expressed in terms of two-parametric
integrals, as at zero temperature, and a sum over $N$ which
takes into account the temperature dependence.

So,  all further steps of calculations necessary to obtain the expression
for the polarization tensor actually coincide with that  in
Ref.\cite{Bordag05}. Within this formalism, the polarization
tensor becomes an expression of the type \be
\label{Pist}\Pi_{\la\la'}(k)=\int_0^\infty\int_0^\infty \ ds \ dt
\ \langle M_{\la\la'}(p,k) \Theta \rangle_T, \ee
where in $M_{\la\la'}(p,k)$ we collected all factors appearing
from the vertexes and from the lines except for that which go into
$\Theta$. By using \Ref{pres} and \Ref{p2res} the average (the
momentum integration over $p$) can be transformed into
\be \langle M_{\la\la'}(p,k) \Theta \rangle_T = M_{\la\la'}(s,t)
\Theta_T  ,\ee
where now $M_{\la\la'}(s,t)$ collects all factors except for
$\Theta_T $.

As in Ref.\cite{Bordag05}, we break the whole polarization tensor
into parts according to the division introduced in
Eq.\Ref{vertexfactor}. We write
\be\label{subdvision}\Pi_{\la\la'}(k)=\sum_{i,j}\Pi^{ij}_{\la\la'}(k)+\Pi^{\rm
ghost }_{\la\la'}(k)+ \Pi_{\la\la'}(k)^{tadpole}\ee
with $i,j=1,2,3$ and
\be\label{}\Pi^{ij}_{\la\la'}(k)=\Gamma^{(i)}_{\mu\nu\la}G_{\mu\mu'}(p)
\Gamma^{(j)}_{\mu'\nu'\la}G_{\nu'\nu}(p-k) \ee
including corresponding subdivisions of $M$.

To organize our calculation, we remind that the PT is calculated
in terms of double parametric integrals in $s$ and $t$. The function
$\langle\Theta (s, t)\rangle$ \Ref{Theta} is symmetric whereas
$\Theta_T (s, t)$ is not, because of the factor $e^{( \frac{i k_4 N
t}{q T} )} = e^{2 (\tilde{u} k)t}$ in the temperature dependent exponential. To
restore the $s \leftrightarrow t$ symmetry, we write
\be\label{texponent} e^{2 (\tilde{u} k)t} = S_T + A_T \ee
where
\be\label{STAT} S_T =\frac{1}{2}\left(e^{(2\tilde{u} k)t}+ e^{2
(\tilde{u} k)s}\right) ,~~~~A_T =\frac{1}{2}\left(e^{2 (\tilde{u} k)t}- e^{2
(\tilde{u} k)s}\right).\ee
Then the function $\Theta_T(s, t)$ can be split  in symmetric and
antisymmetric parts, $\Theta_T = \Theta_T^s + \Theta_T^a$, with
 \bea\label{ThetaSA} \Theta_T^s &=& \langle\Theta(s,
t)\rangle S_T e^{- \frac{N^2}{ 4 T^2 q} },\nn\\
\Theta_T^a &=& \langle\Theta(s, t)\rangle A_T e^{- \frac{N^2}{ 4
T^2 q} }.\eea
With these definitions introduced, the terms entering the
integral in Eq.\Ref{Pist} have the following general structure.
The expressions in $M_{\la\la'}(s,t)$ which are symmetric under $s\leftrightarrow$ go
multiplied by the symmetric temperature factor $\Theta_T^s $, and
 vise versa, the antisymmetric functions go multiplied by
  $\Theta_T^a $. This observation gives us the possibility to make
  use of the results in Ref.\cite{Bordag05},
where the symmetric form factors for the operators $T^{(1)} -
T^{(4)}$ where calculated. At finite temperature, they should be
multiplied by the factor $\Theta_T^s $ in the total expression.
The antisymmetric terms as well as the terms giving contributions
to the remaining form factors must be calculated separately.

In fact, this procedure is rather simple and it is  described in the
previous work. To make the present paper self-contained, we educe
the details of this calculation in the Appendix. In the main text, we
restrict ourselves to the  description of it in general words. For the
first six operators, i.e., for that which do not contain    the vectors $u_\mu$, the terms
appearing after the momentum integrations have the following
forms. First,
\be\label{ts1} \Pi_{\la\la'}=P_\la P^\top_{\la'}
+a\delta^{||}_{\la\la'}+b\delta^{\perp}_{\la\la'}+icF_{\la\la'},
\ee
where $P_\la$ is given in terms of the vectors \Ref{vec},
\be \label{Pla} P_{\la}=r  \ l_\la+\al  \  id_\la +\beta  \
h_\la\ee
and $P^\top_{\la'}$ is the transposed expression, $r,\alpha,
\beta, a , b, c$ are some functions specific for different parts
of $\Pi_{\la\la'}$.

The second type of expressions  has a  slightly more complicated
form \be\label{ts2} \Pi_{\la\la'}=P_\la Q_{\la'}+Q^\top_\la
P^\top_{\la'} +a \delta^{||}_{\la\la'}+b
\delta^\perp_{\la\la'}+icF_{\la\la'},\ee
with $P_\la$ from \Ref{Pla} and \be\label{Qla} Q_\la=s \
l_\la+\gamma \ id_\la+ \delta \ \ h_\la,\ee
where $s, \gamma, \delta$ are some other functions.  Then, from the
requirement that the weak transversality condition
\Ref{doubletransv} holds, one can derive the expressions standing  in front of
the operators $T^{(1)} - T^{(6)}$. This procedure is efficient
but, of course,  not obligate.

The form factors standing in front of the operators $T^{(7)} - T^{(10)}$
can be obtained  after
integration in accordance with Eqs.\Ref{pres}, \Ref{p2res} as that part of the expressions which are proportional to one or to two powers of $u_\mu$. To
make clear the structure of the expressions which appear  after
the averaging procedure, we note that it results in the 
formal substitution of $p_\mu$ in the initial expressions by
$p_\mu \to \tilde{p}_\mu + \tilde{u}_\mu = [(\frac{A}{D}k) +
\tilde{u}]_\mu$ in the final ones. So, the $u$-dependent parts are
easily determined. The necessary details  for different parts of
the polarization tensor are given in the next section.

We complete this section with the description of the
renormalization procedure adopted. As it is well known, the
divergent parts of the polarization tensor do not depend on the
temperature and field. So, each form factor  can be written as
follows
 \be\label{renorm}\Pi^{(i)}(B,T)= [\Pi^{(i)}(B,T) -
 \Pi^{(i)}(B=0,T=0)] + \Pi^{(i)}(B=0,T=0).\ee
Then, the expression in the brackets is finite whereas the last
term is divergent and must be renormalized  by the standard
procedure in quantum field theory. In terms of the resummed series
\Ref{ThetaT1},\Ref{pres},\Ref{p2res} the term with $N=0$ just
corresponds to the zero temperature case. So, actually the above
procedure has relevance to these terms, only. As a result, we
obtain the renormalized polarization tensor at finite temperature
in the field presence which is the object of interest.

\section{Calculation of the form factors}
In this section we calculate the contributions to the form factors
stemming from individual terms $\Gamma^{(i)}_{\mu\nu\la}$ ($i=1,2,3$) introduced in   Eq.\Ref{vertexfactor}. At zero
temperature that was done in Ref.\cite{Bordag05}. So, here we
present mainly that part of calculations having relevance to the
temperature dependence  of these expressions. The first is
\be\label{Pi11}\Pi_{\la\la'}^{11}= (k-2p)_\la
G_{\mu\mu'}(p)(k-2p)_{\la'}G_{\mu'\mu}(p-k)\ee
and it transforms into \bea\label{M11}
&& M_{\la\la'}^{11}(s,t)\Theta_T =\left\langle (k-2p)_\la (k-2p(s))_{\la'}E^{-s}_{\mu\mu'}E^{-t}_{\mu'\mu} \ \Theta \right\rangle_T \nn\\
&&= \left[ \left(\left(1-2\frac{A}{D}\right)k -
2\tilde{u}\right)_{\la} \left(\left(1-2E^{s}\frac{A}{D}\right) k -
2\tilde{u} \right)_{\la'} \nn \right.
\\ && \left.
-4i \left( \frac{E^{-s}F}{D^{\top}} \right)_{\la\la'}  tr E^{-s-t}
 \right]\Theta_T . \eea
We note the property
$E^{s}\frac{A}{D}=\left(\frac{A}{D}\right)^\top$. The trace is
\be\label{expf} tr E^{-s-t} =2 \left[1 + \cosh(2q)\right]. \ee
Remind   the variable $q = s+t$.

Here we also introduce the variable $\xi = s-t$ which is
antisymmetric with respect to the replacement $ s \leftrightarrow t$ and assume $s$ and $t$ to be replaced, $s=(q+\xi)/2$ and $t=(q-\xi)/2$.

 Using the notation of Eq.\Ref{ts1} we define
\bea\label{P11}P_\la&=&\left(\left(1-2\frac{A}{D}\right)k\right)_{\la}\nn\\
&=&\frac{\xi}{q} l_\la+ 2i d_\la \frac{\ss\st}{\sq}+
h_\la \frac{\sxi}{\sq} \nn\\
&\equiv&r  l_\la+i\al d_\la+\beta h_\la \eea
and below we will express the obtained results for different form
factors in terms of  $r, \al, \beta$.

The second part in the expression \Ref{M11} has to be integrated
by parts. This procedure at finite temperature requires a special
explanation.
 We represent
\be\label{ds-dt} \frac{-2iFE^{-s}}{D^{\top}}= \left(\frac{\pa}{\pa
\xi}\right)   \left[ r \delta^{||}-iF\frac{\cxi}{\sq}
+\delta^{\perp}\beta + C \right], \ee
where $C$ is a constant, i.e., it must be independent on $\xi$. We include it in the integration
procedure in order to make use of this parameter, see below.
 The derivative with
respect to $\xi$ will be integrated by parts. We should note that
expressions which are symmetric under an exchange of $s$ and $t$, i.e., which depend on $q$ only, are not affected in Eq.\Ref{ds-dt}. So, in
 $\Theta_T$ we
have to differentiate  only  the terms  which depend on $\xi$,
\be \label{deriv} \frac{\pa}{\pa \xi} \Theta_T = \left(\frac{1}{2}
\ B_1 -  2 (\tilde{u}k) \right)  \Theta_T, \ee
 with the notation
\be \label{B1a}B_1\equiv r l^2+ \beta h^2. \ee
Then, after integration by parts, the last term in the
Eq.\Ref{M11} gives the contribution
\bea\label{partint} -4i \left( \frac{E^{-s}F}{D^{\top}}
\right)_{\la\la'} &=&  2 ( 1 + cosh(2q))(2 (\tilde{u}k) - B_1)\nn
\\ &&\left[ r \delta^{||}-iF\frac{\cxi}{\sq}
+\delta^{\perp}\beta + C \right]\Theta_T, \eea
that has to be added to the first part of the equation. It contains a term $4 \tilde{u}_\mu
\tilde{u}_\nu \Theta_T$
which is  is the quadratic in $\tilde{u}_\mu$. Now we chose the constant $C = -
2\frac{\tilde{u}_\mu\tilde{u}_\nu}{(\tilde{u}k)}$ in a way that 
this term cancels and considerably simplifications in  further calculations appear.   In
this a way we arrive at the final expression at finite
temperature. As concerns the surface term, it is canceled by the
contribution of the tadpole diagram, see below.

Applying formula \Ref{A1} from the Appendix for the function
\Ref{ts1} with the parameters $ a = r B_1$ and $b = \beta B_1$  ,
giving the contributions to the first six form factors, and
gathering  the factors at the $u$-dependent structures giving rise to the operators
$T^{(7)}-T^{(10)}$, we obtain for  $ M^{11}_{\la\la'}$:
 \bea\label{M11b}
M^{11}(s,t)&=& \left\{
 -r^2 T^{(1)} + \left( \al^2 - \beta^2
\right) T^{(2)} - r \beta T^{(3)} \right.
 \nn \\ && \left.
 - \al r T^{(4)} + [ B_1 \coth(q) - 2 (\tilde{u}k)] T^{(6)} \right.
 \nn \\ && \left.
 + 2 (\tilde{u}k)[ - r T^{(7)} - \beta T^{(8)} + \al T^{(9)}] \right\}
 2 ( 1 + \cosh(2q)). \eea
This rather simple expression includes the symmetric   and
antisymmetric  with respect to the $ \leftrightarrow t$ terms.
Since the integral in $s, t$ is symmetric, actually, the factor
$\Theta_T^s$ stands at the former terms, and $\Theta_T^a$ at the
latter ones. This remark concerns all the expressions written
below for other parts of $\Pi_{\la\la'}$.

The next contribution is $M^{22}$. From \Ref{vertexfactor} we get
\bea\label{Pi22}
\Pi^{22}_{\la\la'}&=&4\left(\delta_{\la\mu}k_{\nu}-\delta_{\la\nu}k_{\mu}\right)G_{\mu\mu'}(p)
\left(\delta_{\la'\mu'}k_{\nu'}-\delta_{\la'\nu'}k_{\mu'}\right)G_{\nu'\nu}(p-k)
\eea
and
\bea\label{M22}M^{22}_{\la\la'}(s,t)&=&4\left(
E^{-s}_{\la\la'}(kE^{-t}k)+E^t_{\la\la'}(kE^{-s}k) \right. \\ &&
\left.-\left(E^{-s}k\right)_{\la}\left(E^{-t}k\right)_{\la'}
-\left(E^{t}k\right)_{\la}\left(E^{s}k\right)_{\la'}\right).\nn
\eea
Here we have  expression of the type of \Ref{ts2}. The
parameters are
\be \begin{array}{rclrclrcl}
r&=&1,&\al&=&\sinh(2t),&\beta&=&\cosh(2t),\\
s&=&1,&\gamma&=&\sinh(2s),&\delta&=&\cosh(2s),
\end{array}
\ee and
\beao a&=&-2l^2-h^2(\cosh(2s)+\cosh(2t)),\\
b&=&-l^2(\cosh(2s)+\cosh(2t))-2h^2\cosh(2s)\cosh(2t).\eeao
Using formula  \Ref{A2} from the Appendix we obtain
\bea\label{M22b} M^{22}(s,t)&=&8T^{(1)}+8\cosh(2(s+t))T^{(2)} \nn
\\ && + 4(\cosh(2s)+\cosh(2t))T^{(3)}\nn \\
&& - 4 (\sinh(2s)- \sinh(2t))T^{(4)} . \eea
Note that the last form factor is antisymmetric and does not
contribute at zero temperature.

Now we consider  $\Pi^{12}$ and $\Pi^{21}$. From
\Ref{vertexfactor} we get
\bea\label{Pi12} \Pi^{12}_{\la\la'}+\Pi^{21}_{\la\la'}  &=&
 \delta_{\mu\nu}(k-2p)_{\la}
G_{\mu\mu'}(p)
2\left(\delta_{\la'\mu'}k_{\nu'}-\delta_{\la'\nu'}k_{\mu'}\right)G_{\nu'\nu}(p-k)
\nn\\&&+
2\left(\delta_{\la\mu}k_\nu-\delta_{\la\nu}k_\mu\right)
G_{\mu\mu'}(p)
 \delta_{\mu'\nu'}(k-2p)_{\la'}G_{\nu'\nu}(p-k),\nn
\eea and further
\bea\label{M12a} \left(M^{12}_{\la\la'}+M^{21}_{\la\la'}
\right)\Theta_T   &=&  \left\langle 2\left\{ (k-2p)_{\la}
\left(\left(E^{s+t}-E^{-s-t}\right)k\right)_{\la'} \nn
\right.\right. \\&&\left. \left.
        + \left(\left(E^{s+t}-E^{-s-t}\right)^\top k\right)_{\la} (k-2p(s))_{\la'} \right\}
        \Theta\right\rangle_T .
  \eea
We use the averages \Ref{pres}  and obtain,

\bea\label{M12} \left(M^{12}_{\la\la'}+M^{21}_{\la\la'}
\right)\Theta_T &=& 2\left\{\left(\left(1-2\frac{A}{D}\right)k -
2\tilde{u} \right)_{\la}
 \left(Q k \right)_{\la'} \right.\nn
  \\&& \left.
        + \left(Q^\top k\right)_{\la}
        \left( \left(1-2\frac{A^{\top}}{D^{\top}}
        \right)k -
2\tilde{u}\right)_{\la'} \right\}
     \Theta_T .
 \nn \eea
This is an expression of the form of \Ref{ts2} with $P_\la$ from
Eq.\Ref{P11} and with additional $u$-dependent terms, where

\be\label{Q12} Q\equiv E^{s+t}-E^{-s-t}=2iF\sinh(2q). \ee
Then from Eq.\Ref{A2} and accounting for the $u$-dependent
structures we find
\bea\label{M12b}  M^{12}(s,t)+M^{21}(s,t)&=& 4
\sinh(2q)\Bigg[ - 2 \al T^{(2)}
  + r T^{(4)}    \nn  \\ &&  
  + ( 2 (\tilde{u} k) - B_1 ) T^{(6)} - \frac{i N}{q T}
  T^{(9)}\Bigg].\eea

Next we consider the contribution of $\Pi^{33}$ together with the
contribution  from the ghosts, $\Pi^{\rm ghost}$. We get
\bea\label{Pi33}\Pi^{33}&=& \left(
\delta_{\la\mu}(p-k)_\nu+\delta_{\la\nu}p_{\mu}\right)G_{\mu\mu'}(p)
 \nn \\ && \left(\delta_{\la'\mu'}(p-k)_{\nu'}+\delta_{\la'\nu'}p_{\mu'}\right)G_{\nu'\nu}(p-k).
\eea We use the property of the propagator
\be\label{cp} p_{\mu}G_{\mu\mu'}(p)=G(p)p_{\mu'} \ee
 and obtain
after simple calculation, using, for instance, the cyclic property
of the trace,
\bea\label{Pi33a}\Pi^{33}_{\la\la'}&=&G_{\la\la'}(p)G(p-k)(p-k)^2
+p^2G(p)G_{\la'\la}(p-k) 
\nn\\ &&
+p_{\la}G(p)(p-k)_{\la'}G(p-k)
+(p-k)_{\la}G(p)p_{\la'}G(p-k).\nn \eea
In the first two lines in the r.h.s. one line collapses into a
point by means of, e.g., $p^2G(p)=1$ and the corresponding graph
becomes a tadpole like contribution which will be considered  below separately
together with the tadpole diagram. The last two lines in the
r.h.s. are just equal to the contribution from the ghosts, second
line in \Ref{NPT1}, with opposite sign and cancel. So we obtain
\be\label{Pi33+ghosts}\Pi^{33}+\Pi^{\rm ghost}=0.\ee
Let us turn to the $\Pi^{13}$ and $\Pi^{31}$ contributions. We start
from
\bea\label{Pi13} \Pi^{13}_{\la\la'}&=&
\delta_{\mu\nu}(k-2p)_{\la}G_{\mu\mu'}(p)
\left(\delta_{\la'\mu'}(p-k)_{\nu'}+
\delta_{\la'\nu'}p_{\mu'}\right)G_{\nu'\nu}(p-k) , \nn
\\
\Pi^{31}_{\la\la'}&=&
\left(\delta_{\la\mu}(p-k)_{\nu}+\delta_{\la\nu}p_{\mu}\right)
G_{\mu\mu'}(p) \delta_{\mu'\nu'}(k-2p)_{\la'}G_{\nu'\nu}(p-k)
\eea and arrive at
\beao\label{M13} M^{13}_{\la\la'}(s,t)&=&-\left\langle(k-2p)_{\la}
\left(E^{\rm sy}(k-2p(s))+E^{\rm as}k\right)_{\la'}
\Theta \right\rangle_T ,\nn \\
M^{31}_{\la\la'}(s,t)&=&-\left\langle \left(E^{\rm
sy}(k-2p)-E^{\rm as}k\right)_{\la}(k-2p(s))_{\la'} \Theta
\right\rangle_T, \eeao
where the notation
\bea\label{Eas}
E^{\rm sy}&=&\frac12\left(E^{s+t}+E^{-s-t}\right)=\delta^{||}+
\delta^{\perp}\cosh(2q),\nn \\
E^{\rm
as}&=&\frac12\left(E^{s+t}-E^{-s-t}\right)=iF\sinh(2q)=\frac{1}{2}Q
\eea
is introduced. Remind that the contributions appearing
 after the averaging procedure can be obtained by the substitution
$p_\mu \to \tilde{p}_\mu + \tilde{u}_\mu = [(\frac{A}{D}k) +
\tilde{u}]_\mu$ in the final expressions.

Doing so  we obtain for the averages
\bea\label{M13a}
M^{13}_{\la\la'}(s,t)&=&\left\{-\tilde{P}_{\la}\tilde{Q}_{\la'}+4i\left(\frac{E^{\rm sy}E^{-s}F}{D^\top}\right)_{\la\la'}\right\} \Theta_T \nn ,\\
M^{31}_{\la\la'}(s,t)&=&\left\{-\tilde{Q}^\top_{\la}\tilde{P}^\top_{\la'}+4i\left(\frac{E^{\rm sy}E^{-s}F}{D^\top}\right)_{\la\la'}\right\}\Theta_T, \nn \\
\eea
 where $\tilde{P}_\la$ is given by Eq.\Ref{P11} with the mentioned
replacement being done, and $\tilde{Q}_\la = Q_\la - 2
\tilde{u}_\la$ with
\bea\label{88}
Q_{\la'}&&=\left(\left(E^{\rm as}+E^{\rm
sy}\left(1-2\frac{A}{D}\right)^\top \right) k\right)_{\la'}\nn \\
&&= r l_{\la'}+id_{\la'}\Bigg(\sinh(2q)-\al \cosh(2q)\Bigg)
+\beta\cosh(2q)h_{\la'} \nn\\&&\equiv s\ l_{\la'}+\gamma \
id_{\la'}+ \delta ~ h_{\la'}. \eea
 The second contributions to the both lines in the 
r.h.s. in Eq.\Ref{M13a} must be integrated by parts.

This expression differs from that of in Eq.\Ref{ds-dt} by the
factor $E^{sy}$. That results in the following replacements in the
Eqs.\Ref{ds-dt},\Ref{partint}, $\beta \to \beta \cosh(2q)\equiv \delta,
 \frac{\cosh(\xi)}{\sinh(q)} \to \frac{\cosh(\xi)}{\sinh(q)}
\cosh(2q)\equiv \tilde{\gamma}$. With these substitutions done and
accounting for the overall factor 8, we calculate this
contribution,
 \be\label{partint2}
8 i \frac{E^{sy}E^{-s}F}{D^{\top}} = [2B_1 - 4(\tilde{u}k)](r
 \delta^{||} + \delta \delta^{\perp} - i \tilde{\gamma} F ).
 \ee
 Again, we  have chosen the constant $C = - 2
 \frac{\tilde{u}_\mu \tilde{u}_\nu}{(\tilde{u}k)}$
 and omitted  the surface term. The latter will be considered separately.
The above expression gives the parameters $a, b, c$ entering the
form \Ref{ts2} for this part. By using Eq.\Ref{A2} from Appendix
and the structure of the $u$-dependent tensors, we obtain
\bea\label{Pi13a}&& \Pi^{13}+\Pi^{31}=2 r^2 T^{(1)} \nn \\&& +
2\left\{ [\beta^2- \al^2] \cosh(2q) +
\al \sinh(2q)\right\}T^{(2)} \nn \\
&&+ r \beta \left(1+\cosh(2q)\right) \ T^{(3)} \nn \\
&& - 2 r \sinh(q) \sinh(\xi)  T^{(5)}\nn \\
&&+  \left[  2 (\tilde{u}k) ( \al - \gamma + 2 \tilde{\gamma})
\right.  \nn \\ && \left.  + l^2 r (\gamma -\al -
2\tilde{\gamma})+ h^2\beta (\gamma -\al \cosh(2q) -
2\tilde{\gamma})\right]
 T^{(6)} \nn \\
 && \frac{i N}{q T}\left[ 2 \frac{r}{k_4} T^{(7)} +  \frac{\beta(1 +
 \cosh(2q))}{k_4} T^{(8)} \right. \nn \\ && \left. + r (\al - \gamma)
 T^{(4)} + ( \gamma - \al ) T^{(9)} + k_4 \beta (1 - \cosh(2q)) T^{(10)}
 \right] .
  \eea
Finally, we need $\Pi^{23}$ and $\Pi^{32}$. Proceeding in the same
way as before, we derive from \Ref{NPT1} and \Ref{vertexfactor}
\bea\label{Pi23}
\Pi^{23}_{\la\la'}&=&2(\delta_{\la\mu}k_{\nu}-\delta_{\la\nu}k_{\mu})
G_{\mu\mu'}(p)
(\delta_{\la'\mu'}(p-k)_{\nu'}+\delta_{\la'\nu'}p_{\mu'})G_{\nu'\nu}(p-k) ,\nn \\
\Pi^{32}_{\la\la'}&=&2(\delta_{\la\mu}(p-k)_{\nu}+\delta_{\la\nu}p_{\mu})G_{\mu\mu'}(p)
(\delta_{\la'\mu'}k_{\nu'}-\delta_{\la'\nu'}k_{\mu'})
G_{\nu'\nu}(p-k) ,\eea which gives
\bea\label{M23a} M^{23}_{\la\la'}(s,t)\langle\Theta\rangle &=&
 \bigg\langle 2\bigg\{
-E^{-s}_{\la\la'}\left(kE^{t}(k-p(s))\right)
-E^{t}_{\la\la'}\left(kE^{-s}p(s)\right) \nn \\ &&
+\left(E^{-s}p(s)\right)_{\la}\left(E^{-t}k\right)_{\la'}
+\left(E^{t}(k-p(s))\right)_{\la}\left(E^{s}k\right)_{\la'}
\bigg\} \Theta\bigg  \rangle_T , \nn \\
M^{32}_{\la\la'}(s,t)\langle\Theta\rangle &=&
\bigg\langle 2\bigg\{ -E^{-s}_{\la\la'}\left(kE^{-t}(k-p)\right)
-E^{t}_{\la\la'}\left(pE^{-s}k)\right) \nn \\ &&
+\left(E^{-s}k\right)_{\la}\left(E^{-t}(k-p)\right)_{\la'}
+\left(E^{t}k\right)_{\la}\left(E^{s}p\right)_{\la'} \bigg\}
\Theta\bigg\rangle_T . \eea
Now the average \Ref{pres} must be used. As it was noted, this
results in the substitution $p_\mu \to \tilde{p}_\mu +
\tilde{u}_\mu$ in Eq.\Ref{M23a}. For the operator $\tilde{p}(s)$
the matrix $(\frac{A^{\top}}{D^{\top}}k)$ must be substituted. In
this way the $u$-dependent part is calculated. The $u$-independent
part can be rearranged according to
\be\label{rearr}M^{23}_{\la\la'}+M^{32}_{\la\la'}=2(A+B)\ee
with
\bea\label{A} A&=&-2E^{t}_{\la\la'}\left(k\frac{E^t-1}{D}k\right)
+\left(E^{t}k\right)_{\la}\left(\left(\frac{E^{t}-1}{D}\right)^\top
k\right)_{\la'}
 \nn \\ &&
 +\left( \frac{E^{t}-1}{D}  k\right)_{\la}\left(E^{-t}k\right)_{\la'} , \nn \\
B&=&-2E^{-s}_{\la\la'}\left(k\frac{E^s-1}{D}k\right)
+\left(E^{-s}k\right)_{\la}\left(\frac{E^{s}-1}{D} k\right)_{\la'}
 \nn \\ &&
+\left( \left(\frac{E^{s}-1}{D} \right)^\top
k\right)_{\la}\left(E^{s}k\right)_{\la'}   . \nn \eea $A$ and $B$
have the structure of \Ref{ts2}. For $A$ we obtain
\bea\label{AA1}
r'&=&1,~ \al' = \sinh(2t),~\beta'=\cosh(2t)\nn \\
s'&=&\frac{t}{q},~\gamma'= \frac{1}{2} \al,~\delta'=
\frac{\cs\st}{\sq} \eea
 and
\bea\label{}
a'&&= -2\left( \frac{t}{q}l^2 + \delta' h^2 \right), \nn \\
b'&&= \cosh(2t) a', ~ c' = \sinh(2q) a',
 \eea
 and for $B$
\bea\label{B1}
r''&=&1 ,~ \al''= -\sinh(2s),~ \beta''= \cosh(2s) \nn \\
s''&=&\frac{s}{q},~ \gamma''= ~-\gamma',~ \delta''= \frac{\sinh(s)
\cosh(t)}{\sq} \eea
and
\bea
a''&=&-2\left(\frac{s}{q}l^2+\delta'' h^2\right), \nn \\
b''&=&\cosh(2s)a'' , ~c''= - \sinh(2s)a''. \eea
 For convenience, in the formulas \Ref{AA1} and \Ref{B1} we used apostrophes 
  for denoting similar parameters  having the same structure in different parts of Eq. \Ref{ts2}.

Putting these contributions together we obtain with \Ref{A2}
\bea\label{Pi23a}&&
\Pi^{23}+\Pi^{32}=2\Bigg\{-2T^{(1)} \nn \\
&&-2\Bigg[\frac{\cosh(2t)\cs\st+\cosh(2s)\ss\ct}{\sq} \nn \\ &&
+2\cosh(\xi)\ss\st\Bigg] T^{(2)} \nn \\ &&
-\left[1+\frac{s\cosh(2s)+t\cosh(2t)}{q}\right]  T^{(3)} \nn \\
&&+  \frac{s\sinh(2s)-t\sinh(2t)}{q}T^{(4)} \nn \\
&& +\left[-1+\frac{s\cosh(2s)+t\cosh(2t)}{q}\right] T^{(5)} \nn \\
&& + \left[(\frac{s\sinh(2s)-t\sinh(2t)}{q})l^2 + 2
\cosh(q)\sinh(\xi)
h^2\right] T^{(6)} \nn\\
&&+ 2 (\tilde{u}k) \left[- \sinh(q)\cosh(\xi)T^{(6)} -
\frac{\sinh(q)\sinh(\xi)}{k_4} T^{(8)} \right. \nn \\ && \left. -
\sinh(q)\cosh(\xi) T^{(9)} + \sinh(q)\sinh(\xi) T^{(10)}\right]
 \Bigg\} . \eea

Now we consider the contribution of the tadpole diagram, the last
line in Eq.\Ref{NPT1}. Accounting for the explicit form of the
propagator,
\be\label{prop} G_{\mu\nu}(p) = \int\limits_0^{\infty}d q  e^{- q
p^2} E^{- q}_{\mu\nu}, \ee
and calculating $tr G(p) = 2 \int\limits_0^{\infty}d q  e^{- q
p^2}(1 + \cosh(2q)) $ and the bracket averages, we obtain
\bea\label{Ptp} \Pi^{tp}_{\mu\nu} &&=\left\langle G_{\mu\nu} +
G_{\nu\mu} - 2 tr G \delta_{\mu\nu} \right\rangle_T \nn\\
&&= - 2 \frac{1}{(4\pi)^2}
\sum\limits_{N=-\infty}^{+\infty}\int_{0}^{\infty} d q
e^{(-\frac{N^2}{4 q T^2})} \nn\\
&& \frac{[\delta^{||}_{\mu\nu} (1 + 2 \cosh(2q))+
\delta^{\perp}_{\mu\nu}(2 + \cosh(2q))]}{q \sinh(q)}. \eea
The contribution from the tadpole like terms coming from
$\Pi_{\mu\nu}^{33}$, Eq.\Ref{Pi33} equals just  the first two
terms in the Eq.\Ref{Ptp}. That adds the extra terms
$2(\delta^{||}_{\mu\nu} + \delta^{\perp}_{\mu\nu}\cosh(2q))$ into
the numerator of the above expression and cancels the first and
the last terms in the total. Hence, the final contribution of the
tadpoles and the tadpole like terms is
\be\label{Ptp1} \Pi^{tp}_{tot} = - 4 \frac{1}{(4\pi)^2}
\sum\limits_{N=-\infty}^{+\infty}\int_{0}^{\infty} d q
\ e^{-\frac{N^2}{4 q T^2}} \ \frac{\delta^{||} \cosh(2q)+
\delta^{\perp}}{q \sinh(q)}. \ee

Now let us consider the surface contributions from the sum $
\Pi^{sf} = \Pi^{(11)} + \Pi^{(13)} + \Pi^{(31)}$:
\bea\label{st}\Pi^{sf}=&& 4\int\limits_{surface} ds dt \left[
\delta^{||} \frac{\xi}{q} \cosh(2q)) + \delta^{\perp}
\frac{\sinh(\xi)}{\sinh(q)}\right. \nn \\  && \left. - i F
\frac{\cosh(\xi)}{\sinh(q)} - 2 \frac{\tilde{u}\cdot
\tilde{u}}{(\tilde{u}k)} \cosh(2q) \right] \Theta_T, \eea
where the explicit expressions for the parameters $r, \beta$ from
Eq.\Ref{P11} are substituted. To relate the integrals in $q$ and
$s,t$ we introduce an integration variable $v=\xi/q$ in place of $q$. In this case $\frac{\pa}{
\pa \xi} = \frac{1}{q} \frac{\pa}{ \pa v}$, and
$\int\limits_{0}^{\infty} ds dt \to \frac{1}{2}
\int\limits_{0}^{\infty} q dq \int\limits_{-1}^{+1}dv $. Hence,
the integration over $\xi$ results in an integration by parts  over
$v$. In the new variables the above equation reads,
 \bea\label{st}\Pi^{sf} &&= 2\int\limits_0^{\infty} dq \Theta_T
 \left[ \delta^{||} v \cosh(2q) + \delta^{\perp}
\frac{\sinh(q v)}{\sinh(q)}\right. \nn \\  && \left. - i F
\frac{\cosh(q v)}{\sinh(q)} - 2 \frac{\tilde{u}\cdot
\tilde{u}}{(\tilde{u}k)} \cosh(2q) \right]_{v=-1}^{v=+1}, \eea
with
\be\label{Thetav} \Theta_T = \exp\left[ 2 (\tilde{u}k) \frac{1}{2} q(1
- v) - \frac{N^2}{4 q T^2}\right] \langle \Theta\rangle, \ee
where
\be \langle \Theta\rangle = \frac{1}{(4\pi)^2} \frac{ e^{-k(
\delta^{||} \frac{1}{4}(1-v^2) + \delta^{\perp} \frac{S
T}{S+T})k}}{q \sinh(q)}\ee
and $\frac{S T}{S+T} =
\frac{\tanh[\frac{1}{2}q(1-v)]\tanh[\frac{1}{2}q(1+v)]}{\tanh[\frac{1}{2}q(1
-v)]+\tanh[\frac{1}{2}q(1+v)]}$. Hence it follows that
\bea\label{Thetal} (\Theta_T)^{+1}
&=&\frac{1}{(4\pi)^2}\frac{e^{-\frac{N^2}{4qT^2}}}{q
\sinh(q)},\nn\\
(\Theta_T) _{-1}&=&
\frac{1}{(4\pi)^2}e^{\frac{iNk_4}{T}}\frac{e^{-\frac{N^2}{4qT^2}}}{q
\sinh(q)}. \eea
After substitution of these functions in Eq.\Ref{st} we obtain
\bea\label{st1}\Pi^{sf} &&= 2\int\limits_0^{\infty}dq
\sum\limits_{N=-\infty}^{+\infty} (\Theta_T)^{+1}
 \left[ \left(\delta^{||}\cosh(2q)+ \delta^{\perp}\right)
 \left(1+ e^{\frac{iNk_4}{T}}\right)
\right. \nn \\  && \left. - \left(i F \coth(q)+2
\frac{\tilde{u}\cdot \tilde{u}}{(\tilde{u}k)}\cosh(2q) \right)
\left(1- e^{\frac{iNk_4}{T}}\right) \right],. \eea
Now we note that in the
imaginary time formalism the external momentum $k_4 = 2\pi n_k T$
with $n_k = 0, \pm 1, \pm 2,.. $. The phase factor in the exponential
is $\frac{ N k_4}{T} = 2\pi N n_k$. So, the exponentials in the
brackets equal to 1. This results in the factor 2 for the first
two terms and zero for the third term in the integrand. To find the
last term we also note that for $k_4 \not= 0$ the denominator
$(\tilde{u}k)\not = 0$ and the bracket is zero too. For $k_4 =
0$, we expand the exponential in a series and find $-2 \left(1-
e^{\frac{iNk_4}{T}}\right)\frac{\tilde{u}\cdot
\tilde{u}}{(\tilde{u}k)}|_{k_4=0}= - \frac{N^2}{q T^2}$. Thus, we
come to the conclusion that the last term contributes in the static
limit $k_4 = 0$, only.

As a final step, we take the Eq.\Ref{st1} and the tadpole
contribution Eq.\Ref{Ptp1} together. The terms in front of $\delta^{II}$ and
$\delta^{\perp}$ do cancel in the total. So, the only
contribution is the $u$-dependent part coming from $\Pi^{sf}$. We
denote it as $\Pi^{sf}_{tot}$,
\be\label{sttot} \Pi^{sf}_{tot}= - \frac{4}{(4\pi)^2} \int
\limits_0^{\infty}dq\sum\limits_{N=-\infty}^{+\infty}\frac{\tilde{u}\cdot
\tilde{u}}{(\tilde{u}k)}~ \frac{\cosh(2q)}{q
\sinh(q)} \ e^{-\frac{N^2}{4qT^2}} \left(1-
e^{\frac{iNk_4}{T}}\right). \ee
This term contributes in the static limit only and it is transversal by
itself. This is because  being multiplying by $k_\la$ it is zero
due to the difference in the curly brackets, if $k_4 \not = 0$.
The integrand is nonzero if $k_4 = 0$, but for the product $k_\la
(\Pi^{sf}_{tot})_{\la\la'} = 0$ holds. In fact, it is the limiting expression
of the tensor $T^{(8)}$ for $k_4 = 0$. However, we will consider
this part separately for convenience. Thus, we collect all the
contributions coming from individual parts.

 Now let us gather them together  to obtain the one-loop form
factors $M^{(i)}(s,t)$ for the polarization tensor. We present the
results as the list of explicit functions of variables $q = s+t$
and $\xi = s-t$:
\bea\label{Mia} M_1&=& 4 - 2(\frac{\xi}{q})^2 \cosh(2q,)\nn \\
M_2&=&4 \frac{1 -\cosh(q)\cosh(\xi)}{(\sinh(q))^2} - 2 + 8 \cq\cxi, \nn \\
M_3&=&- 2\cosh(2q) \frac{\xi\sxi}{q\sq} - 2 + 6\cxi\cq, \nn \\
M_4&=&2 \frac{\xi}{q}\left( \sinh(2q) - \frac{\cq -
\cxi}{\sq}\right) - 6 \cq\sxi, \nn \\
M_5&=& - 2 + 2 \cq\cxi, \nn \\
M_6^{(1)}&=& 2\left[\frac{\xi}{q}\coth(q)(1 - 3(\sq)^2) +
\sxi\cq\right]l^2 \nn \\ &+& 2\left[\frac{\sxi}{\sq}\coth(q)(1 -
3(\sq)^2) + 2\sxi\cq \right]h^2 \nn \\
M_6^{(2)}&=& \frac{i N}{q T} k_4~2~ \left(\sinh(2q) -
\coth(q)\right),
\nn \\
M_7&=&\frac{i N}{q T }\frac{1}{k_4}\frac{\xi}{q}(- 2 \cosh(2q)), \nn \\
M_8&=&\frac{i N}{q T} \frac{1}{k_4} \left(-2 \frac{\sxi}{\sq} - 4
\sq\sxi\right), \nn \\
M_9&=& \frac{i N}{q T}2\left[\frac{\cq - \cxi}{\sq} - \sinh(2q) -
2 \sq\cxi \right], \nn \\
M_{10}&=& 0. \eea
The symmetric form factors have to be multiplied by $\Theta_T^s$
and the antisymmetric ones - by $\Theta_T^a$, when the integration
over $s,t$ is carried out.  It is interesting that $M_{10}$ is
zero in one-loop order.

Thus, according to Eq.\Ref{exp}, we presented the polarization
tensor in the form
 \bea\label{NPBT} \Pi_{\la\la'}(k)&=&
 \sum\limits_{i=1}^{9}T_{\la\la'}^{(i)}\Pi^{(i)}(k) +
 (\Pi^{sf}_{tot})_{\la\la'}, \nn \\
\Pi^{(i)}(k)&=&\sum_{N=-\infty}^{+\infty}\int\limits_0^{\infty}
 d s d t ~ M^{(i)}(s,t)\Theta_T \eea
 as double parametric integrals over the proper time parameters $s,t$
 and the temperature sum. The last term in the fist line is written in
 Eq.\Ref{sttot}. This representation is crucial for what
 follows. It preserves as much symmetries of the polarization tensor as possible.

The obtained expression for the polarization tensor could be used
in various applications. In the next three sections we consider
the zero field limit, the Debye mass  and the magnetic mass of
gluons in the magnetic background field at high temperature.
\section{Limit of zero background field}
To make a link of our formalism with a usual one, let us consider
the limit of zero background field, $B=0$. In our dimension less variables
this simply corresponds to the limit $q, \xi$ go to zero. In this
case, the most form factors and operators also go to zero. More
precise, we have for form factors,
 \be\label{M1-3B0a} M_1 = M_2 = M_3 = 4 -
 2\left(\frac{\xi}{q}\right)^2,\ee
 and
\be\label{M1-3B0b} M_7 = M_8 =  \frac{i N}{q
T}\frac{1}{k_4}\left(- 2 \frac{\xi}{q}\right).\ee
All the other operators or form factors are zero. In accordance
with Eqs.\Ref{sumTip},\Ref{sum78} this means that we obtain two transversal operators with the correct formfactors at finite temperature
\cite{Kalashnikov84}. Hence, at $B=0$, the polarization tensor can be written as
\bea\label{PtB0}\Pi_{\la\la'}(k,T&=&\frac{1}{(4\pi)^2}
\sum\limits_{N=-\infty}^{+\infty}\int_{0}^{\infty} ds dt
\left[\frac{e^{-k^2\frac{s t}{q}}}{q^2}~ e^{(2(\tilde{u}k) t -
\frac{N^2}{4 q T^2})}\right.\nn \\ && \left.  \left( 4 -
 2\left(\frac{\xi}{q}\right)^2\right) K_{\la\la'} - 2 \left(\frac{i N}{q
T}\frac{1}{k_4} \frac{\xi}{q}\right)B_{\la\la'} \right] \nn \\
&& -\frac{4}{(4\pi)^2}\sum\limits_{N=-\infty}^{+\infty} \int
\limits_0^{\infty}dq\frac{\tilde{u}_\la
\tilde{u}_{\la'}}{(\tilde{u}k)}~
\frac{e^{-\frac{N^2}{4qT^2}}}{q^2} \left(1-
e^{\frac{iNk_4}{T}}\right), \eea
where we  substituted the surface contribution Eq.\Ref{sttot} at
$B=0$.

In the representation in terms of a series resummed according to Ref. \Ref{resum}, the value
$N=0$  corresponds to the zero temperature case. In Eq.\Ref{PtB0}
this is
\be\label{Pt00}\Pi_{\la\la'}^{(T=0)}=\frac{1}{(4\pi)^2}
\int_{0}^{\infty} ds dt \frac{e^{-k^2\frac{s t}{q}}}{q^2}  \left(
4 -
 2\left(\frac{\xi}{q}\right)^2\right) K_{\la\la'}\ee
that coincides with Eq.(103) in Ref.\cite{Bordag05}. This part
must be renormalized in a standard way. Actually, normalizing to $B=0$ at $T=0$ it must be simplify skipped.

Now, let us calculate the Debye mass squared defined as the limit
of the form, $m^2_{D} = - \Pi_{44}(T,k_4=0, \vec{k} \to
0)$\cite{Kalashnikov84}. Within the representation \Ref{PtB0},
only the last term contributes and we obtain
\be\label{md0}m^2_{D}=\frac{1}{4\pi^2}\sum\limits_{N=1}^{+\infty}
\frac{N^2}{T^2} \int \limits_0^{\infty}dq
\frac{e^{-\frac{N^2}{4qT^2}}}{q^3}.\ee
The integral is simply calculated, $\int \limits_0^{\infty}dq
\frac{e^{-\frac{N^2}{4qT^2}}}{q^3} = \frac{16 T^4}{N^4}$, and the
sum is expressed through Riemann's Zeta-function, $\zeta(2) =
\frac{\pi^2}{6}$. Thus, we obtain
 $m^2_{D}= \frac{2}{3}T^2$, which  is the well known result \cite{Kalashnikov84}.

The next important parameter is the  "magnetic" mass squared which
can be determined as the limit for transversal with respect to the
external field  direction components, $m_{magn}^2 = -
\Pi_{12}(T,k_4=0, \vec{k} \to 0)$. In this case, the operator
$K_{12}$ in Eq.\Ref{PtB0} contributes (the component $B_{12} =
0)$. To calculate the form factor in the high temperature limit,
it is convenient to make an inverse resummation according to
Eq.\Ref{resum}. In the static limit, $k_4 = 0$, the parameter $a =
0$ and we have
 \be\label{resum0} \sum\limits_{N=-\infty}^{+\infty}
e^{( - \frac{N^2}{4 q T^2})}= 2\pi T
\left(\frac{q}{\pi}\right)^{1/2}
\sum\limits_{N=-\infty}^{+\infty}e^{(- 4\pi^2 T^2 N^2 q)}. \ee
For the components $\Pi_{12}$ we then get
\bea\label{Pi12a}&&\Pi_{12}(k,T)=\frac{1}{(4\pi)^2}
\sum\limits_{N=-\infty}^{+\infty}\int\limits_{0}^{\infty} \frac{d
q}{q} \int\limits_0^1 d u (2\pi T)
\left(\frac{q}{\pi}\right)^{1/2} \nn\\
&& e^{(-k^2 q u(1-u)})~ e^{(- N^2 4 \pi^2 T^2 q)} \left( 4 -
 2\left(\frac{\xi}{q}\right)^2\right)~ K_{1 2},
 \eea
where new variables, $s,t \to s=q u, t = q (1 - u)$, were
introduced. The high temperature limit corresponds to the value of
$N=0$ in the Matsubara sum. In that case the integrals can be  
calculated easily. First we compute the integral over $q$ which delivers  $\Gamma$-function, $\Gamma(\frac{1}{2})/(k^2 u(1
- u))^{1/2}$. Then the integration over $u$ gives $3 \pi$. So, for
the form factor we obtain $\Pi(k)_{12}^{(1)} =
\frac{3}{8}\frac{T}{k}$. Hence for the Green function,
\be\label{GfB0}G^0_{12} = \left(-\frac{k_1 k_2}{k^2}\right)
\frac{1}{k^2 - \frac{3}{8} k T}\ee
follows.
Here $k$ is the length of the three-momentum vector $\vec{k}$.
This expression has a fictitious pole that was an old problem of
gauge theories at finite temperature \cite{Kalashnikov84}.

From the above  calculations we see how to proceed  in the present
formalism. The procedure remains actually the same when the field
is switched on. Bellow, we investigate an influence of the field
on the Debye mass and the existence of the fictitious pole.
\section{Debye mass in the presence of the background field }
The gluon Debye mass squared in the background field is defined as before,
$m^2_D = - \Pi_{44}(T,B,k_4=0, \vec{k} \to 0)$, where we have to
use the expression Eq.\Ref{sttot} from the polarization tensor. In
this limit it reads
\be\label{mDB} m_D^2(B) = \frac{1}{4\pi^2} \int
\limits_0^{\infty}\frac{dq}{q}\sum\limits_{N=1}^{+\infty}
\frac{N^2}{ qT^2} ~ \frac{B \cosh(2Bq)}{ \sinh(B
q)}e^{-\frac{N^2}{4qT^2}}, \ee
where the dimensional parameters are restored. We investigate
the case for the field and temperature ratio $s = \frac{B}{4 T^2}
\ll 1$ and  present Eq.\Ref{mDB} in the form: $m^2_D =
\frac{2}{3}T^2 f(s)$, where   the function $f(s)$ is to be
computed. It satisfies the condition $f(0) = 1$. With this
parameter introduced we return again to the dimension less
variable $B q \to q$ and write $f(s)$ in the form, %
\be\label{fs} f(s) = \frac{6}{\pi^2} s^2 \int
\limits_0^{\infty}\frac{dq}{q^2}\sum\limits_{N=1}^{+\infty} N^2
\left[\frac{1}{q} + \frac{ \cosh(2q)}{ \sinh( q)} -
\frac{1}{q}\right]~ e^{-\frac{N^2s}{q}}, \ee
where the first term in the square bracket delivers the zero field limit $f(0) = 1$.
To calculate the other terms  resummations will be done. By
differentiating Eq.\Ref{resum} with respect to $Z$ with $a = 0, Z
= \frac{s}{q}$, we rewrite $f(s)$ as follows
\bea\label{fs1} f(s) = &&1+ \frac{3}{2}\frac{1}{\pi^2} \sqrt{\pi
s}\int
\limits_0^{\infty}\frac{dq}{\sqrt{q}}\sum\limits_{N=-\infty}^{+\infty}
\left[ e^q - e^{-q}  \right. \nn \\&& \left. +
\left(\frac{1}{\sinh(q)} - \frac{1}{q} \right)\right]
 \left( 1 - 2 \frac{\pi^2 N^2 q}{s}\right) e^{(-\frac{\pi^2 N^2
q}{s})}. \eea
Here we made a rearrangement of the integrand in the above
equation and split it into three parts -  the tachyonic one $f_t(s)$
coming from $e^q$, $f_2(s)$, and $f_3(s)$ coming from the curly
brackets. Then we calculate the leading terms of the high
temperature expansion which is given by the terms with $N=0$ in
the temperature sum, next to leading - by the terms with $N\ge 1$.

Now, let us calculate contributions from $N=0$. For $f_t(s,N=0)$ we
get
 \be\label{ft0}f_t(s,N=0) = \frac{3}{2}\frac{1}{\pi^2} \sqrt{\pi
s}\int \limits_0^{\infty}\frac{dq}{\sqrt{q}} \ e^q \ee
This integral diverges at the upper limit that reflects the
tachyonic mode in the spectrum of charged gluons \Ref{spectr}. To
obtain its value we  make the inverse Wick rotation in the
$q$-plain, that is, replace $q \to q e^{-i\pi}$. After
that a simple integration yields
\be\label{ft0a}f_t(s,N=0)= - i\frac{3}{2}\frac{1}{\pi^2} \sqrt{\pi
s} \sqrt{\pi} = - i \frac{3}{2\pi} \sqrt{s}.\ee
The second term is
 \be\label{f2}f_2(s,N=0) =- \frac{3}{2}\frac{1}{\pi^2} \sqrt{\pi
s}\int \limits_0^{\infty}\frac{dq}{\sqrt{q}} \ e^{-q} \ee
and can be easily computed, $f_2(s,N=0) = - \frac{3}{2\pi}
\sqrt{s}$. The third term is
 \be\label{f30}f_3(s,N=0) = \frac{3}{2}\frac{1}{\pi^2} \sqrt{\pi
s}\int \limits_0^{\infty}\frac{dq}{\sqrt{q}}
\left(\frac{1}{\sinh(q)} - \frac{1}{q}\right) ,\ee
that can be expressed trough a Zeta- function, $f_3(s,N=0) =
\frac{3}{2}\frac{1}{\pi^2} \sqrt{\pi s}\sqrt{2\pi}
\zeta(\frac{1}{2},\frac{1}{2})$. Finally we derive
\be\label{f30a} f_3(s,N=0) = \frac{3}{\sqrt{2}\pi}\left(\sqrt{2}-1\right)
\zeta\left(\frac{1}{2}\right) \sqrt{ s}.\ee
Similar simple integrations for the value $N=1$ result in
 \be\label{f1} f(s,N=1) = \frac{25}{4} \frac{\zeta(3)}{\pi^4} s^2.
 \ee
Thus, for the function $f(s) $ we obtain
\bea\label{fsr} f(s) = &1& + \left[
\frac{3}{\sqrt{2}\pi}\left(\sqrt{2}-1\right) \zeta\left(\frac{1}{2}\right) -
\frac{3}{2\pi} \right] \sqrt{ s} \nn \\ &+& \frac{25}{4}
\frac{\zeta(3)}{\pi^4} s^2 - i \frac{3}{2\pi} \sqrt{s}.\eea
Hence, for the Debye mass we derive
\bea\label{mDBfin} m^2_D(B) &&= \frac{2}{3}T^2 \left[1 - 0.8859
\left(\frac{\sqrt{B}}{2T}\right) + 0.4775 \left(\frac{B^2}{16 T^4
}\right) \right.\nn \\
&&\left. - i~ 0.4775 \left(\frac{\sqrt{B}}{2T}\right) +
O(\frac{B^3}{T^6})\right], \eea
where the numeric values of the coefficients in Eq.\Ref{fsr} are
substituted. This expression is interesting in two respects. In
presence of the field, the screening mass for color Coulomb
forces is decreased as compared to the zero field case. This
behavior was already determined in Ref.\cite{Bordag00} in
calculation of other type. The imaginary part of $m^2_D$ is
signaling the instability of the states because of the tachyonic
mode in the spectrum \Ref{spectr}. However, the numeric value of
it is small as compared to the real one. It is of order of the
usual plasmon damping factor at finite temperature.
\section{Transversal modes in the field presence}
Let us investigate the behaviour of the transversal modes in the
field at high temperature. We have to calculate the mean value of
the polarization tensor in the states given by Eq.\Ref{qedstates} for the
polarizations $s = 1$ and $s = 2$ in the limit of $k_4 = 0,
\vec{k} \to 0.$ Accounting for Eqs.\Ref{Tn},\Ref{TnT},
\Ref{qedstates}, we derive for the mean values
\be\label{Pi11}\langle s=1| \Pi(k) |s=1\rangle = h^2 \Pi_2 + l^2
(\Pi_3 - \Pi_5), \ee
\bea\label{Pi22}\langle s=2| \Pi(k) |s=2\rangle &=&
\frac{h^2~l_4^2}{k^2} \Pi_1 + \left( h^2 + \frac{l^2 + h^2}{k^2}
l^2_3 \right) \Pi_3 \nn \\
&+& \frac{h^4 - l^2 l^2_3}{k^2}\Pi_5, \eea
where $l_4 = k_4, k^2 = h^2 + l_3^2, l^2 = l^2_3 + l^2_4$. To
consider the behaviour of the static modes in the perpendicular
with respect to the field plane we put $l_3 = 0$ and $k_4 = 0$ and
 get
\bea\label{Pitr}\langle s=1| \Pi(k) |s=1\rangle &=& h^2 \Pi_2, \nn
\\
\langle s=2| \Pi(k) |s=2\rangle &=& h^2 \left( \Pi_3 + \Pi_5
\right). \eea
We have to calculate the form factors $\Pi_2, \Pi_3$ and $\Pi_5$
educed in Eqs.\Ref{Mia},\Ref{NPBT} also for this case.

The procedure of calculations is quite similar to that of in the
previous sections. We describe its steps by computing the form
factor $\Pi_5$. To carry out integration over  $\xi$, we change
variables, $s = q u, t = q( 1 - u)$, as in section 6. In the limit
of interest, $h^2 \to 0$, we restrict ourselves to the leading in
this parameter term when the function $\langle \Theta \rangle$
Eq.\Ref{Theta} is substituted.  More precise, we take into account
the first term in the expansion,
 \be\label{Theta0} \langle \Theta\rangle = \frac{1}{(4\pi)^2}
 \frac{1 + O(h^2) }{q \sinh(q)}, \ee
because, being substituted into Eq.\Ref{Pitr}, the $\sim O(h^2)$
terms result in      a next to leading correction. Then we make
resummation of the series in $N$ and take into consideration the
term with $N=0$, that gives the leading high temperature
contribution. In this limit we obtain for the form factors
\be \Pi_i^{(N=0)} = \frac{1}{(4\pi)^{3/2}}
\frac{T}{\sqrt{B}} \int\limits_0^1 d u \int\limits_0^{\infty}
\frac{d q \sqrt{q}}{\sinh(q)} M_i(q,u),
\label{ffi} 
\ee
where  the functions from Eq.\Ref{Mia} should be substituted. Here
we restored the overall  factor $\frac{T}{\sqrt{B}}$. All other
variables are dimension less.

For the function $M_5$, integration over $u$ results in the
expression %
\be\label{M5} M_5(q) = - 2 + 2 \frac{\cosh(q)\sinh(q)}{q} \ee
which after substitution in Eq.\Ref{ffi} leads to the integral
\be\label{Pi5} I_5 = \int\limits_0^{\infty} d q \left[ - 2
\frac{\sqrt{q}}{\sinh(q)} + \frac{e^{- q}}{\sqrt{q}}  + \frac{e^{
q}}{\sqrt{q}} \right]. \ee
The second and the third terms are calculated in
Eqs.\Ref{ft0}-\Ref{f2} in section 7, $I_5^{(2)}= \sqrt{\pi}$,
$I_5^{(3)}= - i \sqrt{\pi}$, and for the first one we compute
$I_5^{(1)} = \frac{1}{2}(-4 + \sqrt{2}) ~\sqrt{\pi}~
\zeta(\frac{3}{2}) $. Thus, we obtain in the total,
\be\label{Pi5a} \Pi_5 =  \frac{1}{(4\pi)^{3/2}} \frac{T}{\sqrt{B}}
\left[- 4.21405 - 1.77245 i \right], \ee
where numeric values of the integrals are substituted.

Similar calculations carried out for the form factors $\Pi_2$ and
$\Pi_3$ yield
\bea\label{Pi23}\Pi_2 &=& \frac{1}{(4\pi)^{3/2}}
\frac{T}{\sqrt{B}} \left[- 5.79894 - 7.08982 i \right],\nn \\
\Pi_3 &=& \frac{1}{(4\pi)^{3/2}} \frac{T}{\sqrt{B}} \left[1.04427
- 8.86227 i \right]. \eea
The above expressions have to be used in Eq.\Ref{Pitr} to obtain
final result. The sum of $\Pi_3 + \Pi_5 $ equals, $\Pi_3 + \Pi_5
= \frac{T}{\sqrt{B}} \left[-3.16978 - 10.6347 i \right]$. The
imaginary part is signaling the instability of the state because
of the tachyonic mode presenting in the vacuum, and the real part
is responsible for the screening of transversal gluon fields.

As concerns the imaginary part. Since its origin is clear,  one
should conclude that the one-loop result needs in additional terms
coming from some resummation at finite temperature in the sector
of charged components similar to that of in
Refs.\cite{Bordag00},\cite{Strelchenko04}. That we left for the
future.

Let us turn to the real part and substitute it in the
Schwinger-Dyson equation
\be\label{SDe} D^{-1}(k^2) = k^2 - \Pi(k) \ee
for the neural gluon Green function. We obtain for the mean values
\bea\label{SDes1}\langle ~s=1~|D^{-1}(h^2)|~s=1 ~\rangle &=& h^2 -
Re
( \Pi_2) ~h^2 \nn \\
&=& h^2 \left( 1 +  5.79894 \frac{T}{\sqrt{B}} \right) \eea
and
\bea\label{SDes2}\langle~ s=2~|D^{-1}(h^2)|~s=2~ \rangle &=& h^2 -
Re
( \Pi_3 + \Pi_5) ~h^2 \nn \\
&=& h^2 \left( 1 + 3.16978 \frac{T}{\sqrt{B}} \right). \eea
These are the expressions of interest.

 Two important conclusions follow from Eqs.\Ref{SDes1},\Ref{SDes2}.
First, for the transversal modes in the field presence,  there is
no fictitious pole similar to that of in Eq.\Ref{GfB0}. Second,
there is no the magnetic screening mass in one-loop order. The
transversal components of gluon field remain long range in this
approximation, as at zero external field \cite{Kalashnikov84}.
These conclusions are in a complete correspondence with the
results obtained in Ref.\cite{Strelchenko05} in the Fujikawa
gauge. In this paper only the static limit $k_4=0$ was
investigated. As it has been discovered in
Refs.\cite{Bordag00},\cite{Strelchenko04}, for the charged
components the magnetic mass in the color magnetic field at high
temperature is generated in one-loop order. So, there is an
essential difference in this point. It is also interesting that
the non transversal tensor structure $T^{(5)}$ contributes to the
transversal state $s = 2$.
\section{Discussion}
In the present paper the operator structure of the neutral gluon
polarization tensor in a constant Abelian chromomagnetic field
at finite temperature is derived. There are ten operator
structures contributing in general case. 
We mention that these structures do not depend on the gauge group; for a SU(3) these are the same as for the SU(2) considered here.
%
As investigated in details in Refs.
\cite{Bordag05},\cite{Bordag06} at zero
temperature,  the polarization tensor is not
transversal. The tensors $T^{(5)}, T^{(6)}, T^{(9)}$ and
$T^{(10)}$ satisfy the weak transversality condition
Eq.\Ref{doubletransv}. As actual one-loop calculations showed, the
form factors $\Pi_5, \Pi_6, \Pi_9$ are nonzero and $\Pi_{10} = 0$.
So, only three non transversal operators contribute in the total
in this approximation.

As at zero temperature Ref.\cite{Bordag05}, to carry out
integration over internal momentum of diagrams in the field
presence which is necessary in calculation of form factors, the
Schwinger algebraic method \cite{Schwinger73} was applied. It has
been extended to account for finite  temperature. As a result, we
have derived the explicit expressions for integrals which present
the form factors in terms of two parametric integrals over the
proper-time parameters $s, t$, as at zero temperature, and extra
series in discrete variable $N$ accounting for the temperature
dependence. This representation preserves gauge invariance and
even Lorentz covariance at each step of calculations as far as possible. It is
simple and convenient for investigations.

Within this representation,  the one-loop form factors for the
polarization tensor of neutral gluons have been obtained in the
background Lorentz-Feynman gauge. Note here that in previous
papers \cite{Nielsen78}, \cite{Borisov84},\cite{Strelchenko05}
this tensor in the field without and with temperature included has
been computed in the Fujikawa gauge. In this non-linear gauge the
structure of the neutral polarization tensor is simpler as
compared to the Lorentz gauge and the tensor is transversal, but
the charged sector has a much more complicated structure.

The tensors Eqs.\Ref{Tn},\Ref{TnT} and form factors
\Ref{Mia},\Ref{NPBT} present the structure of the neutral
polarization tensor off shell and can be used in various
applications. First of all, the spectra of gluon modes in the
field at high temperature can be investigated. They may also serve
as the basis for different kinds of resummations of perturbation
series. This is  because to make a  resummation one needs in the
Green functions of shell. As far as we know, the operator
structure of the gluon tensor in the considered background was not
educed  before. To relate our approach with a common  one at
finite temperature, we have investigated in short the zero field
case and recalculated the corresponding  well known  results.

 As applications, in the present paper we restricted ourselves to the
consideration of two important parameters - the Debye mass of
neutral gluons, Eq.\Ref{mDBfin}, and the magnetic mass of
transversal neutral gluons, Eqs.\Ref{SDes1},\Ref{SDes2}, in the
field at high temperature, in one-loop approximation. It was found
that all the asymptotic expansions for the high temperature limit
are expressed in terms of Riemann's $\zeta$-function. As it
follows from Eq.\Ref{mDBfin}, the screening temperature mass of
plasmons is decreased as compared to the zero field case. That
increases a radius of color Coulomb forces when the background
magnetic field is present. This behavior may be important for
quark-gluon plasma. As concerns the magnetic mass, from
Eqs.\Ref{SDes1},\Ref{SDes2} it follows that it is zero in one-loop
order, as at zero field. At the same time, the fictitious pole,
appearing in one-loop order at $B = 0$ (see Eq.\Ref{GfB0}),
disappears  in the field presence. This is important property of
hot gluon plasma. Already earlier one believed that the elimination of the
fictitious pole requires a resummations of perturbation series
\cite{Kalashnikov84},\cite{Rebhan03}  and it is removed by the
gluon magnetic mass.

In connection with the last result we would like to speculate a
little. If one takes  the one-loop result seriously and the
magnetic mass is zero, that means that neutral gluon fields remain
long range like Abelian  magnetic fields at high temperature. This
is in contrast to the charged gluon fields which acquire the
magnetic mass in the field at high temperature in one-loop order
\cite{Bordag00},\cite{Strelchenko05} and therefore are screened at
long distances. This behaviour is important either for problems of
quark gluon plasma or the early universe because at high
temperature the stable Abelian magnetic fields are spontaneously
created, as it has been derived in the daisy resummations in Refs.
\cite{Enqvist94},\cite{Starinets94},\cite{Bordag00},\cite{Strelchenko04}.

To make a final conclusion about this phenomenon, one has to apply
the super daisy resummations which can be realized on the base of
solution of the Schwinger-Dyson equations for the common system of
the  neutral and charged gluon Green's functions. The general
structure of the Green function for neutral gluon fields  is
derived in the present paper. The case of the charged gluon field
will be considered separately elsewhere.

\section*{Acknowledgement}
One of us (V.S.) was supported by DFG under grant number 436 UKR 17/24/05.  Also he thanks the Institute for Theoretical Physics of Leipzig University for kind hospitality.

\section*{Appendix}
 \renewcommand{\theequation}{A.\arabic{equation}}\setcounter{equation}{0}
 In this appendix we collect  formulas which are used to identify the
contributions to the form factors, i.e., the contributions which
go with the tensor structures $T^{(i)}_{\mu\nu}$.

In the course of calculation, from the graph (Fig.
\ref{figure:Pineu}) contributions appear which have up to a
constant the following structure. First,
\be\label{ts1} {\cal F}^{1}_{\la\la'}=P_\la P^\top_{\la'}
+a\delta^{||}_{\la\la'}+b\delta^{\perp}_{\la\la'}+icF_{\la\la'},
\ee
where $P_\la$ is given in terms of the vectors \Ref{vec},
\be \label{Pla} P_{\la}=r  \ l_\la+\al  \  id_\la +\beta  \
h_\la\ee
and $r$, $\al$ and $\beta$ are some functions of the variables $s$
and $t$. The transposition in $P^\top_{\la'}$ changes the sign of
$d_{\la'}$, $P_{\la'}^\top=r  \ l_\la-\al  \  id_\la +\beta  \
h_\la$.

It can be  seen that the expression in Eq.\Ref{exp} fulfills
\Ref{doubletransv} if $\left(rl^2+\beta h^2\right)^2+al^2+bh^2=0$
holds. In that case it can be represented in terms of form factors
according to
\bea\label{A1}&&{\cal
F}^{1}_{\la\la'}=-r^2T^{(1)}_{\la\la'}+(\al^2-\beta^2)T^{(2)}_{\la\la'}
-r\beta T^{(3)}_{\la\la'}-r\al T^{(4)}_{\la\la'}\nn\\&&
+\frac{r(rl^2+\beta h^2)+a}{h^2}T^{(5)}_{\la\la'}+(ral^2+\al\beta
h^2+c)T^{(6)}_{\la\la'} \eea
which can be checked  by inserting the explicit expressions
\Ref{}.

A second type of expressions appears which has a  slightly more
complicated form, \be\label{ts2} {\cal F}^{2}_{\la\la'}=P_\la
Q_{\la'}+Q^\top_\la P^\top_{\la'} +a \delta^{||}_{\la\la'}+b
\delta^\perp_{\la\la'}+icF_{\la\la'}\ee
with $P_\la$ from \Ref{Pla} and \be\label{Qla} Q_\la=s \ l_\la
+\gamma \ id_\la+ \delta \ \ h_\la,\ee
and $s$, $\gamma$ and $\delta$ are also some functions of the
variables $s$ and $t$. In parallel to the above case, if for
\Ref{ts2} the condition \Ref{doubletransv} is fulfilled   than
$(a+2rsl^2)l^2+(b+2\beta\delta h^2)h^2+(r\delta+s\beta)2l^2h^2=0$
must hold. In that case  the representation in terms of form
factors is \bea\label{A2}{\cal
F}^{2}_{\la\la'}&=&-2rsT^{(1)}_{\la\la'}-2(\beta\delta+\al\gamma)
T^{(2)}_{\la\la'} -(r\delta+s\beta)
T^{(3)}_{\la\la'}\nn\\&&+(r\gamma-s\al) T^{(4)}_{\la\la'}
+\left((a+2rsl^2)\frac{1}{h^2}+r\delta+s\beta\right)
T^{(5)}_{\la\la'} \nn\\&&
+\left(c-(r\gamma-s\alpha)l^2+(\al\delta-\beta\gamma)h^2\right)T^{(6)}_{\la\
\la'} .\eea Formulas \Ref{A1} and \Ref{A2} are used in the section
5 for calculation of form factors.

\end{document}